\documentclass[a4paper,12pt]{article}
\usepackage{amssymb}
\usepackage{amsmath}
\usepackage{amsbsy}
\begin{document}
\title{Bose-Einstein correlations in multiple particle production — from 1959 to 1989}
\author{K.Zalewski \thanks{Partly supported by the MEiN grant P03B04529(2005-2008)}
\\ M.Smoluchowski Institute of Physics
\\ Jagellonian University, Cracow\footnote{Address: Reymonta 4, 30 059 Krakow,
Poland, e-mail: zalewski@th.if.uj.edu.pl}
\\ and\\ Institute of Nuclear Physics, Cracow}

\maketitle
\begin{abstract}
The evolution of the experimental knowledge and of the theoretical ideas about
Bose-Einstein correlation in multiple particle production processes, during the
first thirty years from their discovery, is reviewed.
\end{abstract}

\noindent PACS numbers 25.75.Gz, 13.65.+i \\Bose-Einstein
correlations. \vspace*{1cm}
 \newpage
 \tableofcontents

\section{Introduction}

The Bose-Einstein correlations in multiple particle production processes were
discovered almost fifty years ago \cite{GFG}, \cite{GGL}. Since then many old
results got forgotten. For instance now most people believe that what is being
observed in multiple particle production processes is the Hanbury Brown and
Twiss effect, in spite of the very clear proof to the contrary given by Kopylov
and Podgoretsky \cite{KP4} (cf. \cite{LED2} for a more recent brief discussion
of this point). Concepts are renamed. E.g. what is now called Bose-Einstein or
HBT correlations used to be the GGLP effect. Results are rediscovered and
priorities are ascribed at random. It seems, therefore, useful to summarize and
review critically the old results. We chose the first thirty years, which
corresponds to little more than two hundred papers. Even so, this is a vast
subject and I will be grateful for comments and remarks, which could help to
improve the text. After 1989 the explosion in the number of publications on the
subject has been such that now the number of papers can be estimated to be over
one thousand. Around 1990 the first big review papers devoted to Bose-Einstein
correlations appeared \cite{POD2}, \cite{LOR}, \cite{BGJ}. For obvious reasons,
however, they concentrated on the problems which the authors considered topical
at the time of writing and did not have the hindsight we have now.

 This review is chronological, but not historical. A historical review would
have to include all the wrong and/or sterile ideas which influenced the
research workers in the field. No effort in this direction has been made.

When reviewing the work of many people, done over a long period of
time, the choice of notation becomes a problem. We have chosen a
simple, uniform notation without trying to keep track of the
notations of the various authors. Our notation will be introduced
when necessary, but here we summarize the main points. The
four-momentum of a particle will be denoted $p$ or $p'$ and its
positions in space-time $x$ or $x'$. If the particle is on its
mass shell, the time component of $p$ is $E_p = \sqrt{\textbf{p}^2
+ m^2}$. For a pair of particles we introduce

\begin{eqnarray}\label{}
  K &=& \frac{1}{2}(p_1 + p_2),\quad K' = \frac{1}{2}(p'_1 +
  p'_2),\quad q  =  p_1 - p_2, \quad q' = p'_1 - p'_2,\\  q^2 &=& -Q^2 \\
X &=& \frac{1}{2}(x_1 + x_2),\quad X' = \frac{1}{2}(x'_1 +
x'_2),\quad Y  =  x_1 -
  x_2,\quad Y'  =  x'_1 - x'_2,
\end{eqnarray}
with the obvious notation $\sqrt{Q^2} = Q$. For single particles

\begin{equation}\label{}
  K_1 = \frac{1}{2}(p_1 + p'_1),\quad q'_1 = p_1 -
  p'_1,\quad X_1  = \frac{1}{2}(x_1 + x'_1), \quad
   Y_1  =  x_1 - x'_1.
\end{equation}
The subscript is changed from $1$ to $2$, if the position in space-time and the
four-momentum of the particle are $x_2,p_2$. Sometimes, when only one particle
is considered and there is no risk of confusion with the previous formulae, the
subscripts will be dropped.

Let us note for further reference two useful identities

\begin{eqnarray}\label{eleide}
  p_1x_1 + p_2x_2 &=& 2KX - \frac{1}{2}qY,\\
  p_1x_1 - p_2x_2 &=& KY + qX.
\end{eqnarray}
and their analogues for the primed and subscripted parameters as
well as for the space vectors.

We will use pseudo density matrices
$\tilde{\rho}(\textbf{p}_1,\ldots,\textbf{p}_n;
\textbf{p}_1',\ldots,\textbf{p}_n')$. Their diagonal elements
yield the momentum distributions, but the relations are model
dependent and in general more complicated than for the standard
density matrices. The pseudo density matrices are hardly ever
normalized to unity. Moreover, in the GGLP model discussed in the
next chapter their diagonal elements have to be integrated over
phase space, with energy momentum conservation imposed, in order
to give the momentum distributions (see text).

The radius $R$ of the interaction region means different things in
different papers. For a Gaussian distribution of sources

\begin{equation}\label{}
\rho(\textbf{r})= \sqrt{\frac{\alpha}{\pi R^2}}e^{-\alpha
\textbf{r}^2/R^2}
\end{equation}
$\sqrt{\langle r^2\rangle} = \sqrt{3/2\alpha} R$, but both $\alpha
= 1$ and $\alpha = 1/2$ are being used. For sources distributed on
the surface of a sphere $\sqrt{\langle r^2\rangle} = R$, while if
the sources are distributed uniformly over this sphere
$\sqrt{\langle r^2\rangle} = \sqrt{3/5}R$. One should also keep in
mind that the average of the square of the distance between two
points is $2\langle r^2\rangle$, while e.g. $\sqrt{\langle x^2
\rangle} = \sqrt{1/3}\sqrt{\langle r^2\rangle} $. These
ambiguities have been discussed and illustrated by many examples
from the literature in ref. \cite{BAK}.

In the next section we review the two papers which created the field. Then we
present the subsequent results divided in ten years periods. Finally in the
last section we summarize the highlights of this development.

\section{Beginnings}

\subsection{Discovery}

Bose-Einstein correlations in multiple particle production were discovered
accidentally \cite{GOL} as a byproduct of an unsuccessful attempt to find the
$\rho$ meson, which had just been predicted by  Frazer and Fulco \cite{FRF}.
The group of the Goldhabers at Berkeley was studying annihilations of
antiprotons of momentum 1.05 GeV/c in a propane bubble chamber. Since the
predicted $\rho$ meson had isospin one, they hoped to find a bump in the
invariant mass distribution of the unlike-sign pairs ($\pi^+\pi^-$), and no
bump in the mass distributions of the like-sign pairs ($\pi^\pm\pi^\pm$). They
analyzed the $2\pi^+2\pi^-$ and $3\pi^+3\pi^-$ semi-inclusive
channels\footnote{i.e. channels with an unspecified number of $\pi^0$-s}
\cite{GFG}, but instead of finding the expected bumps discovered that the
distributions of the angle between pion momenta, as measured in the
$p\overline{p}$ cms frame, strongly depended on the pion charges. For like-sign
pion pairs the distributions in $\cos\theta_{\pi\pi}$ were almost flat, while
for unlike-sign pairs  large opening angles were more probable as expected from
momentum conservation,. In fact, the distributions for the unlike-sign pions
were even more peaked than required by phase space, so that the combined
distributions of $\cos\theta_{\pi\pi}$ for all the pion pairs at given
multiplicity agreed within errors with calculations from the statistical model
with Lorentz invariant phase space. In order to characterize the distributions
of the opening angle, the group introduced for each distribution a parameter
which was being very popular for many years,

\begin{equation}\label{}
  \gamma = \frac{n_>}{n_<},
\end{equation}
where $n_>$ ($n_<$) is the number of pion pairs with the c.m.s. opening angle greater
(smaller) than $90^\circ$. Thus, the experimental result was that $\gamma^{like} <
\gamma^{unlike}$ with the value of $\gamma$ calculated from Lorentz invariant phase
space falling in between.

\subsection{The GGLP paper}

It took over a month and the help of an eminent theorist to understand that
what was being observed were correlations due to Bose-Einstein statistics
\cite{GOL}. The resulting paper written by the Goldhabers, Lee and Pais
\cite{GGL} (further quoted GGLP) became a standard reference. According to the
SPIRES data base, which omits citations previous to year 1973, this paper has
been quoted over 300 times.  For many years the effect was being referred to as
the GGLP effect. The realization that an analogous effect had been discovered
earlier in astronomy by Hanbury Brown and Twiss \cite{HBT} and applied to the
determination of the angular radii of stars came only much later \cite{GKP},
\cite{GOL}.

The starting point for GGLP was the statistical model with Lorentz invariant phase
space \cite{SRS}. According to this model the distribution of the particle momenta for
a final state produced in a $p\overline{p}$ annihilation, containing $n$ pions and
nothing else, is given by the formula

\begin{equation}\label{LIPS}
  E_{p_1}\ldots E_{p_n}\frac{dN}{d^3p_1\ldots d^3p_n} = C(n_+,n_0,n_-) \frac{a^n}{n!}
  \delta^4(P - \sum_{i=1}^n p_i).
\end{equation}
Here $a$ is a constant, $P$ is the initial total momentum four-vector and the
momentum four-vectors of the final particles are $(E_{p_i},\textbf{p}_i)$,
$i=1,\ldots,n$. The factor $C$ contains a normalization constant and an isospin
factor dependent on the numbers of positive, neutral and negative pions
($n_+,n_0,n_-$). It will not be discussed here any further. Today it is well
known that prescription (\ref{LIPS}) is very unrealistic, but at the time of
GGLP, when beam energies were low and for comparison with experiment the
formula was integrated over all the degrees of freedom except at most one,
agreement with data was usually reasonably good.

The point made by GGLP was that formula (\ref{LIPS}) contains no
correlations between the electric charges of the pions and their
momenta. Consequently, the distribution of opening angles for pion
pairs, as predicted by the model, does not depend on the charge of
the pair, which is in violent contradiction with the data
\cite{GFG}. They proposed to replace the factor $1/n!$ by a
product of three factors, one for pions of each charge. In view of
future generalizations we will present their argument using
operators and matrices instead of the wave functions they used. By
analogy with the formulae for the density matrices in the momentum
representation, we are looking for a replacement

\begin{equation}\label{}
  1/n! \rightarrow
  \prod_k\tilde{\rho}_k(\textbf{p}_1,\ldots,\textbf{p}_{n_k};\textbf{p}_1,\ldots,\textbf{p}_{n_k}),
\end{equation}
where the product is over the pion charges $k = +,0,-$ and the tildes are there
to remind that $\tilde{\rho}_k$ do not have all the properties of textbook
density matrices. We will call the matrices $\tilde{\rho}$ pseudo density
matrices. This includes a variety of matrices similar to, but not identical
with, the density matrices.

Let us define a single particle density operator

\begin{equation}\label{densin}
  \hat{\rho} = \int d^3x |\textbf{x}\rangle \rho(\textbf{x}) \langle \textbf{x}|
\end{equation}
with the usual normalization condition

\begin{equation}\label{}
  \int d^3x \rho(\textbf{x}) = 1.
\end{equation}
Function $\rho(\textbf{x})$ must be real, because operator
$\hat{\rho}$ is hermitian. This single particle state is an
incoherent superposition of pure states, each corresponding to a
given production point $\textbf{x}$. GGLP considered two weight
functions $\rho(\textbf{x})$. Following the statistical model they
put $\rho(\textbf{x}) = $const within a spherical volume of radius
$R_S$ and zero outside. They also introduced a "Gaussian-shaped
volume" meaning that $\rho(\textbf{x}) \sim
\exp[-\textbf{x}^2/2\lambda]$. Comparing the matrix elements
$\tilde{\rho}(\textbf{p}_1,\textbf{p}_2;\textbf{p}_1,\textbf{p}_2)$
calculated within their model (see further) for the two choices of
$\rho(\textbf{x})$ they found that the results were similar, with
the closest agreement for

\begin{equation}\label{rholam}
  R_S = 2.15\sqrt{\lambda}.
\end{equation}
Therefore, they chose the Gaussian as it leads to simpler
calculations. The density matrix in momentum representation, which
corresponds to (\ref{densin}), is

\begin{equation}\label{desipp}
\rho(\mathbf{p};\mathbf{p}') \equiv \langle
\mathbf{p}|\hat{\rho}|\mathbf{p}'\rangle \sim \int d^3x
\rho(\mathbf{x})e^{-i\mathbf{q}\mathbf{x}} \equiv \left\langle
  e^{i\mathbf{q}\mathbf{x}}\right\rangle.
\end{equation}
The corresponding momentum distribution can be obtained by putting $\textbf{q}
= 0$ and is constant, as was to be expected from the uncertainty principle. Let
us note that $\rho(\textbf{p};\textbf{p}')$ is a single particle density
matrix, but not necessarily the one describing the system being considered.

Let us define an unsymmetrized $n$-particle density matrix as a
product of the single particle density matrices (\ref{desipp})

\begin{equation}\label{denpro}
  \rho^U(\textbf{p}_1,\ldots,\textbf{p}_n,\textbf{p}'_1,\ldots,\textbf{p}'_n) =
  \prod_{i=1}^n \rho(\textbf{p}_i,\textbf{p}'_i).
\end{equation}
Taking the diagonal elements and imposing energy momentum conservation one
could "derive" in this way the statistical model (\ref{LIPS}). GGLP, however,
proposed to symmetrize for each charge of pions. For $k=+,0,-$ their
assumption, written in a more general form \cite{KAR}, is

\begin{equation}
  \tilde{\rho}_k(\textbf{p}_1,\ldots,\textbf{p}_{n_k},\textbf{p}'_1,\ldots,\textbf{p}'_{n_k}) =
   \frac{1}{n_k!}\sum_{P,Q}
  \rho_k^U(\textbf{p}_{P1},\ldots,\textbf{p}_{Pn_k},\textbf{p}'_{Q1},\ldots,\textbf{p}_{Qn_k}'),
\end{equation}
where the summation extends over the $n_k!$ permutations $P$ of
the momenta $\textbf{p}_i$ and over the $n_k!$ permutations $Q$ of
the momenta $\textbf{p}'_i$. This formula can be rewritten as

\begin{equation}\label{symkar}
  \tilde{\rho}_k(\textbf{p}_1,\ldots,\textbf{p}_{n_k},\textbf{p}'_1,\ldots,\textbf{p}'_{n_k}) =
   \sum_{Q}
  \rho_k^U(\textbf{p}_1,\ldots,\textbf{p}_{n_k},\textbf{p}'_{Q1},\ldots,\textbf{p}_{Qn_k}'),
\end{equation}

GGLP considered no more than two pions of the same sign, because they assumed
in the calculations that every event with more than one $\pi^0$ contains
exactly two $\pi^0$-s. Moreover, they needed only the distributions of momenta
i.e. the diagonal elements of the pseudo density matrices. Thus, their recipe
was as follows. For $n_k=0,1$ there is no dependence of the factor
$\tilde{\rho}_k$ on the momenta of the pions. For $n_k = 2$

\begin{equation}\label{rhosym}
  \tilde{\rho}_k(\textbf{p}_1,\textbf{p}_2;\textbf{p}_1,\textbf{p}_2) =
  \tilde{\rho}(\textbf{p}_1;\textbf{p}_1)\tilde{\rho}(\textbf{p}_2;\textbf{p}_2) +
  |\tilde{\rho}(\textbf{p}_1;\textbf{p}_2)|^2,
\end{equation}
where the hermiticity of the single particle pseudo density matrix
has been used. Substituting expression (\ref{desipp}) one finds:

\begin{equation}\label{rhoint}
\tilde{\rho}_k(\textbf{p}_1,\textbf{p}_2;\textbf{p}_1,\textbf{p}_2) = 1 + \left| \int
d^3x \rho(\textbf{x})e^{i(\textbf{p}_2-\textbf{p}_1)\cdot\textbf{x}}\right|^2,
\end{equation}
which for a Gaussian $\rho(\textbf{\textbf{x}})$ gives

\begin{equation}\label{}
\tilde{\rho}_k(\textbf{p}_1,\textbf{p}_2;\textbf{p}_1,\textbf{p}_2) = 1 +
e^{-\lambda(\textbf{p}_1 - \textbf{p}_2)^2}.
\end{equation}
As seen from the derivation, the momenta in the exponent are
three-momenta. GGLP, however, replaced them by four-momenta, in
order to simplify further the calculations. This makes the formula
Lorentz invariant and does not change much their numerical
results. Incidentally, it improved agreement with experiment. Let
us follow GGLP and introduce their "correlation function"

\begin{equation}\label{gglcor}
  \psi(p_1,p_2) = 1 + e^{-\lambda Q^2},
\end{equation}
where $Q^2 = -(p_1 - p_2)^2$. This formula has later become very popular. Let us
stress, however, that in the GGLP model it was a factor in the integrand of the phase
space integral, while later it was being used without further integrations.  For the
channels $2\pi^+2\pi^-$, $2\pi^+2\pi^-\pi^0$ and $2\pi^+2\pi^-2\pi^0$ the GGLP
proposal amounts to replacing in formula (\ref{LIPS}) the factor $1/n!$ by

\begin{equation}\label{}
  \psi(p_1,p_2)\psi(p_3,p_4)\tilde{\psi}(p_5,p_6),
\end{equation}
where $p_1$ and $p_2$ are the four-momenta of the positive pions,
$p_3$ and $p_4$ are the four-momenta of the negative pions and
$\tilde\psi(p_5,p_6) = 1$ unless $n_0 = 2$. For $n_0 = 2$:
$\tilde{\psi}(p_5,p_6) = \psi(p_5,p_6)$, where $p_5$ and $p_6$ are
the four-momenta of the two neutral pions. The results, after
averaging with suitable weights over $n_0$, were found to be in
good qualitative agreement with the corresponding experimental
results from \cite{GFG}. Also the radius of the interaction region
$R_S$ came out between $0.7$ fm and $1$ fm which is the expected
order of magnitude. Quantitatively, however, $\gamma^{unlike}$ was
predicted too small and $\gamma^{like}$ too large whatever the
choice of $R_S$. Thus the model underestimated the difference
between the like- and unlike-sign pion pairs.

Besides their calculations GGLP made some important general
remarks. They realized that "an adequate model should at the same
time give a reasonable account of all combined aspects of the
annihilation process" and that, therefore, one should look for
other evidence for or against the model. They predicted that for
any given exclusive channel the effects of Bose-Einstein
statistics should decrease with increasing energy. They worried
about the multiplicity distributions, because the volume
$\frac{4}{3}\pi R_S^3$, which they found, was significantly
smaller than the volume which in the statistical model governs the
multiplicity distribution. They stressed that the study of
Bose-Einstein correlations may give valuable information about the
reaction mechanism, but that "results of this study should not be
construed  to imply that dynamical effects (such as, for example,
$\pi$-$\pi$ interactions) are definitely negligible". In fact they
suspected that the "$\pi$ isobars", or in today's language
resonances, may be important.

\section{First decade (1961 --- 1970)}

In the sixties the study of Bose-Einstein correlations was gradually gaining
popularity. The conceptual framework was still that of the GGLP paper, though
an isolated attempt to study three- and four-body correlations \cite{FIL}
should be noted. Usually, people plotted the distributions of
cos$\theta_{\pi\pi}$, calculated the parameters $\gamma$, or some other
parameters which could be expressed in terms of the parameters $\gamma$,  and
supported, or very rarely criticized, the implications of the GGLP paper. The
effect found in \cite{GFG} was confirmed for other $\overline{p}p$
interactions. A compilation from 1967 \cite{CZS} lists $\gamma^u,\gamma^l$
pairs for 25 different (exclusive) channels and/or energies. The effect was
found in $\pi^+ p$ \cite{BAR2}, $\pi^-p$ \cite{JUN}, $pp$ \cite{HOL} and $K^+p$
\cite{BAE} interactions. The effective attraction in momentum space was found
for $\pi^0\pi^0$ pairs \cite{ESK}, but not for $K^+\pi^\pm$ pairs \cite{BAE}.
The results for the $\pi^0$-s and the $K^+\pi^+$ pairs eliminated the
possibility \cite{RUB} that the effect for $\pi^\pm\pi^\pm$ pairs was due to
the fact that these pairs are "exotic" i.e. unable to form resonances, and not
to the fact that they consist of identical bosons. A related possibility was
that the effect is due to resonances. Here the results were not so clear (cf.
eg. the review \cite{ZAL1} and references given there), however, cutting out
the resonances did not change much the measured $\gamma$-s \cite{BAE}.
Moreover, a particularly strong effect was observed for $\pi^+ p$ interactions
at $8$ GeV/c with nine charged pions and a proton or ten charged pions and a
neutron in the final state, where the energy per particle is too low for a
significant production of resonances \cite{ESW}. Thus Bose-Einstein
correlations remained as the only plausible explanation.

Some qualitative implications of the GGLP model got confirmed. A compilation of the
data for annihilations showed \cite{CZS} that with increasing energy the effect, as
measured by the parameter

\begin{equation}\label{}
  C = \frac{\gamma^u - \gamma^l}{(\gamma^u + 1)(\gamma^l + 1)},
\end{equation}
decreases rapidly. A rough fit was \cite{CZS}

\begin{equation}\label{}
  C = \frac{73.9}{n_c^{1.64} (n_0+1)^{1.54}}e^{- \frac{\sqrt{s}}{0.132(n_c+n_0)}},
\end{equation}
where $n_c$ and  $n_0$ denote the numbers of charged and neutral pions
respectively. The centre of mass energy $\sqrt{s}$ is here expressed in GeV.
The uncertainty of the normalizing factor $73.9$ was about fifty percent, while
the uncertainties of the other three parameters were at the ten per cent level.
For events with many particles in the final state it was possible to show that
the effect occurs in every event and not only in some special ones \cite{ESW}.
The effect is enhanced when one chooses pairs of pions with similar values of
momenta $|\textbf{p}_i|$ i.e. with similar energies \cite{BAR2}.

Quantitatively, however, the model did not work. The conclusion in \cite{CZS}
was that "The model of Goldhaber et al. is in violent disagreement with the
data". The main argument was that at low energy per particle the values
predicted for $C$ are much too small. Moreover, assuming a constant interaction
radius, $R = 1.3$ fm, the authors found that for the exclusive process
$\overline{p}p \rightarrow 3\pi^+3\pi^-$  the model grossly underestimates the
energy dependence of parameter $C$. Incidentally, \cite{CZS} seems to have been
the first calculation with three identical particles in the final state. The
statistical model contains no peripherality and consequently with increasing
energy per particle its predictions for the single particle distributions
become completely wrong. The GGLP model did nothing to correct that, but the
point was that also its predictions for the correlations were wrong. One tried
to improve the starting point by replacing the statistical model by the
uncorrelated jet model, which includes peripherality, or by the C{\L}A model,
which included moreover some multiperipheral correlations, but this did not
help \cite{BAE}. The situation was summarized at the 1-st multiparticle
conference in Paris \cite{ZAL1}. To put it short: the GGLP model was good
enough to convince people that the effect is due to Bose-Einstein statistics,
but not good enough to make quantitative predictions.

\section{Progress in the seventies}

\subsection{The approaches of Kopylov and Podgoretskii, and of Shuryak}

In 1971 an important series of papers by Kopylov, Podgoretskii and
collaborators (further quoted KP) begun to appear \cite{GKP}. The approach of
KP differed significantly from that of GGLP. There was a major difference in
motivation. The purpose of GGLP was to explain the charge dependence of the
distributions of the opening angles for pion pairs. They noticed that the
radius of the interaction region, obtained as a byproduct, was of the expected
order of magnitude, though probably too small, but this remark was not
followed. According to a detailed review \cite{BGJ} only two experimental
papers previous to 1976 gave values for the radius of the interaction region
(\cite{CZS} and \cite{DON}). Looking up these papers, one sees that in both
cases the radii were given only to convince the reader that the authors worked
hard before recognizing that they are unable to fit the data.  KP were familiar
with the work of Hanbury Brown and Twiss \cite{HBT} and their purpose was to
use quantum interference, for identical fermions as well as for identical
bosons, to get information about the lifetimes and extensions of various
sources. They considered a variety of sources \cite{KP2}: resonances of
particles and of nuclei, highly excited nuclei where both lifetimes and shapes
were of interest and interaction regions in multiple particle production
processes. KP did not compare their results with experimental data.

 The GGLP approach did not include time. It corresponds to a picture where all the pions
are emitted instantaneously and simultaneously. Then, when it exactly happened
is irrelevant.  The sources of KP were time dependent and extended in time.
They considered it to be the main difference between the two approaches
\cite{GKP} and claimed to have "a more correct theoretical approach"
\cite{KP2}.  Another difference was technical, but of great practical
importance. GGLP had reasonably simple formulae for the integrands of the phase
space integrals. The integration over phase space, however, remained to be
done. In practice this limited their calculations of the physical distributions
to exclusive channels with no more than six particles. KP assumed that the
constraints of energy and momentum conservation have little effect on the
distributions of the momentum difference $(q_0, \textbf{q}) = (E_{p_1} -
E_{p_2}, \textbf{p}_1 - \textbf{p}_2)$in the region of small $|\textbf{q}|$.
Thus, their results could be directly compared with the measured
cross-sections. This eliminated the necessity of evaluating the phase space
integrals and made the predictions applicable also to inclusive processes.
Obviously, it was a reasonable assumption only for sufficiently high
multiplicities for exclusive processes, or for sufficiently high energies for
inclusive processes. The KP results for multiple particle production were
sketched in \cite{KP1} and then described in more detail in \cite{KP2}. In ref.
\cite{KP3} the theory was reformulated using a more powerful formalism borrowed
from optics and some generalizations were given. A short summary, including
important new ideas, was given in \cite{KOP2}. Reference \cite{KP4} explained
the relation between the work of GGLP and that of Hanbury Brown and Twiss.

Both GGLP and KP considered an incoherent superposition of pure pion states produced
by sources excited at some common time, say $t = 0$, at various points of space
$\textbf{x}$ and emitting one pion per source. The GGLP sources, however, emit the
pions instantly and without changing their locations. The KP sources are much more
general.  The amplitude of a pion emitted by a source was assumed to be the solution
of the Klein-Gordon equation

\begin{equation}\label{}
  \left(\nabla^2 - \frac{\partial^2}{\partial t^2} - m^2\right)A(x) = -4\pi J(x),
\end{equation}
where $m$ is the pion mass, $x$ is the four-vector $\textbf{x},t$
and $J(x)$ is the (classical) source. Further there will be many
sources contributing incoherently, but let us start with just one.
The equation simplifies when the Fourier transforms in time:

\begin{equation}\label{}
  A(\textbf{x};\varepsilon) = \int_{-\infty}^\infty dt A(x)e^{i\varepsilon t};\qquad
  J(\textbf{x};\varepsilon) = \int_{-\infty}^\infty dt J(x)e^{i\varepsilon
  t};
\end{equation}
are introduced. Note that, according to the general rules of
quantum mechanics, $\varepsilon$ can be interpreted as the energy
of the pion. The transforms satisfy the equation

\begin{equation}\label{klegor}
 \left(\nabla^2 - \textbf{p}^2\right)A(\textbf{x};\varepsilon) =
 -4\pi J(\textbf{x};\varepsilon),
\end{equation}
where $\textbf{p}^2 = \varepsilon^2 - m^2$ and, therefore, it is
the square of the momentum of the pion. The solution of this
equation is well known and described in detail in \cite{KP3}, but
we will need here only a special case. Let us assume that all the
source functions $J(x)$ are negligible outside a small region of
space $V$ and that the detector is at a large distance $r_D$ from
the centre of region $V$. Large means here much larger than the
diameter of region $V$. Then, in the vicinity of the detector, the
(approximate) solution is

\begin{equation}\label{}
  A(x;\varepsilon) = \tilde{A}(p)\frac{e^{i|\textbf{p}|r_D}}{r_D};\qquad \tilde{A}(p) =
  \int d^4x J(x)e^{ipx};
\end{equation}
where $p$ is the pion energy-momentum four-vector $(\textbf{p},E_p)$ with the
momentum $\textbf{p}$ oriented from the centre of region $V$ to the detector
and $|\textbf{p}| = \sqrt{\textbf{p}^2}$. Note that all the dependence on the
source is contained in the factor $\tilde{A}(p)$. The other factors are
irrelevant from the point of view of symmetrization and will be further
omitted. Thus in general one source, say the one labelled $i$,  contributes to
the single particle pseudo density matrix a term
$\tilde{A}_i(p)\tilde{A}_i^*(p')$. In general this product depends on some
parameters and has to be averaged over them. The result should be summed over
all the sources. Note that the contributions from all the space-time points
where $J(x)$ is significantly different from zero are summed coherently (adding
amplitudes) in the formula above, while the averaging and the summation over
sources correspond to incoherent (adding probabilities) superpositions. Finally

\begin{equation}\label{shurya}
  \tilde{\rho}(\textbf{p},\textbf{p}') = \sum_i \left\langle \tilde{A}_i(p)\tilde{A}_i^*(p')\right\rangle.
\end{equation}
Formulae of this kind, which are more general than the KP model,
were introduced by Shuryak \cite{SHU1, SHU2}. Shuryak's results
can be derived starting from the density operator

\begin{equation}\label{shuevo}
  \hat{\rho}(t) = \int_{-\infty}^tdt_0\int_{-\infty}^tdt'_0
  e^{-iH(t - t_0)}|J(t_0)\rangle\langle J(t'_0)|e^{+iH(t - t'_0)}.
\end{equation}
Actually it is a pseudo density operator, because its trace does
not have to be equal one. Further we will take $t \rightarrow
\infty$. When $J(x) = \langle \textbf{x}| J(t)\rangle$ is a
classical current this leads to the KP theory. $J(x)$, however,
can be any single particle production amplitude (cf. e.g.
\cite{PCZ}). This formula, as it stands, is for one coherent
source. In order to make a realistic model it is necessary to sum
over the sources and, sometimes, to average over some parameters
of the sources. Including that, introducing, after the first and
before the last factor of the integrand, unit operators built from
the eigenstates of the position operator and converting to a
pseudo density matrix element in the momentum representation we
find

\begin{equation}\label{rhoshu}
  \tilde{\rho}(p,p') = \int d^4x \int d^4x'\langle \textbf{p}|e^{-iH(t -
  t_0)}|\textbf{x}\rangle \sum_i \left\langle J_i(x)J_i^*(x')\right\rangle
  \langle \textbf{x'}|e^{iH(t - t'_0)}|\textbf{p}'\rangle.
\end{equation}
Evaluating the matrix elements of the time-evolution operators and
omitting the uninteresting factor $\exp[it(E_{p'} - E_p)]$ we get

\begin{equation}\label{}
 \tilde{\rho}(p,p') = \int d^4x \int d^4x'  \left\langle \sum_i J_i(x)J_i^*(x')\right\rangle
  e^{i(px-p'x')},
\end{equation}
or equivalently in other variables

\begin{equation}\label{shkqxy}
\tilde{\rho}(p_1,p'_1) = \int d^4X_1 \int d^4Y_1  \sum_i \left\langle J_i(X_1 +
\frac{Y_1}{2})J_i^*(X_1 - \frac{Y_1}{2})\right\rangle e^{i(q_1X_1 + K_1Y_1)},
\end{equation}
This formula is the starting point for much modern work.

KP proposed

\begin{equation}\label{}
  J(x) = \tau^{-1/2}\delta(\textbf{x} - \textbf{x}_0 -\textbf{v}t) e^{-iEt - \frac{t}{2\tau}}\theta(t).
\end{equation}
This corresponds to a classical point source created at time $t=0$ at point
$\textbf{x} = \textbf{x}_0$ and moving from that point on with constant
velocity $\textbf{v}$. The source has energy $E$ and decays emitting a pion
according to an exponential decay law with average life time $\tau$. Thus the
parameters $\textbf{x}_0, \textbf{v}, E, \tau$ depend on the source label $i$.
The integrations necessary to get $\tilde{A}(p)$ are in this case trivial and
one finds using (\ref{shurya})

\begin{equation}\label{}
  \tilde{\rho}(p,p') = \sum_i \left\langle \frac{\tau e^{-i\textbf{q}\cdot \textbf{x}_0}}
  {\left(\frac{1}{2} + i\tau(E -\varepsilon +\textbf{p}\cdot \textbf{v})\right)
  \left(\frac{1}{2} - i\tau(E - \varepsilon' + \textbf{p'}\cdot \textbf{v})\right)}\right\rangle.
\end{equation}
Here KP assumed that the averaging should be done over the source energy $E$
and that this can be reduced to an integration over $E$ from minus to plus
infinity. The integral is easily done using the method of residues. The sources
are labelled by the points $\textbf{x}_0$ of their creation. Thus, the
summation over $i$ reduces to an integration over space with some weight
$\rho(\mathbf{x}_0)$. Finally, omitting an irrelevant factor $2\pi$,

\begin{equation}\label{kpdema}
  \tilde{\rho}(\textbf{p},\textbf{p}') = \frac{\int d^3x_0 \rho(\textbf{x}_0)e^{-i\textbf{q}\cdot \textbf{x}_0}}
  {1 - i\tau(q_0 - \textbf{q}\cdot \textbf{v})}.
\end{equation}
Usually one puts $\textbf{v} = 0$ \cite{KOP2} so that

\begin{equation}\label{}
|\tilde{\rho}(\textbf{p},\textbf{p}')|^2 = \frac{|\int d^3x_0
\rho(\textbf{x}_0)e^{-i\textbf{q}\cdot \textbf{x}_0}|^2}{1 + \tau^2q_0^2}.
\end{equation}
The symmetrization is done as in GGLP. Thus for the two-body correlations formula
(\ref{rhosym}) remains valid. Since $\tilde{\rho}(\textbf{p},\textbf{p}) = 1$, the
first (non interference) term in (\ref{rhosym}) remains equal one. The second
(interference) term is modified by the factor $(1 + \tau^2 q_0^2)^{-1}$. From the
experimental point of view this is a beautiful formula. From the dependence of the
correlation function on the vector $\textbf{q}$ at fixed $q_0$ one can determine the
space distribution of the sources just like in the GGLP model. Studying the dependence
on $q_0$ at fixed $\textbf{q}$, however, one can additionally measure the life time of
the source.

An obvious objection to formula (\ref{kpdema}) is that a density matrix depending on
the difference of the momenta only is grossly unrealistic. GGLP did not have this
difficulty, because they kept only the states allowed by energy-momentum conservation
and the necessary projection introduces a dependence on $p_1 + p_2$ . KP \cite{KP3}
proposed the following way out. Suppose that the incoherent sources are not point-like
in space, but smeared. Thus

\begin{equation}\label{kproex}
  \hat{\rho} = \int d^4x_s e^{-iH_0(t-t_s)}|\psi_s\rangle \rho(x_s)
  \langle \psi_s|e^{+iH_0(t-t_s)}.
\end{equation}
Here the four-vector $x_s$ labels the sources. It gives the space and time
position of the source at the moment when the source got created, or of any
other point in space-time which defines unambiguously the source. Function
$\rho(x_s)$ gives the distribution in space-time of the incoherent sources,
thus, for any function $f(x_s)$ the average over the sources is

\begin{equation}\label{}
\left\langle  f(x_s) \right\rangle \equiv \int d^4x_s \rho(x_s)f(x_s).
\end{equation}
$H_0$ is the free particle Hamiltonian. The exponentials give the time dependence of
the state vectors $|\psi_s\rangle$ and their dependence on the time component of
$x_s$. The state vectors $|\psi_s\rangle$ are equivalent in the sense that

\begin{equation}\label{}
\langle \textbf{x}|\psi_s\rangle = \psi(\textbf{x} - \textbf{x}_s).
\end{equation}
Thus all the incoherent sources are related by rigid shifts in
space-time. It is convenient to use a picture where the density
matrix does not depend on the time $t$. This can be achieved
either by using the time dependent momentum eigenstates
$e^{-iH_0t}|\textbf{p}\rangle$, or equivalently by using the
interaction picture, where each state vector gets an additional
factor $e^{+iH_0t}$, which cancels the $t$-dependent factors both
in the density operator and in the time dependent momentum
eigenstates. Let us define the function

\begin{equation}\label{}
  A(\textbf{p}) = \int d^3x e^{-i\textbf{px}}\psi(\textbf{x}).
\end{equation}
Then, taking the matrix element of the density operator and dropping some constant
factors,

\begin{equation}\label{rhokop}
  \tilde{\rho}(\textbf{p}_1,\textbf{p}_2) =
  A(\textbf{p}_1)A^*(\textbf{p}_2)\left\langle e^{iqx_s} \right\rangle.
\end{equation}
This formula has several interesting features. The diagonal elements, which yield the
single particle momentum distribution are

\begin{equation}\label{}
\tilde{\rho}(\textbf{p},\textbf{p}) = |A(\textbf{p})|^2.
\end{equation}
Thus, any single particle distribution can be reproduced perfectly. Function
$A(\textbf{p})$ is related to the properties of a single source defined by the
state vector $|\psi\rangle$, but tells us nothing about the distribution of
sources in space-time. On the other hand, for the correlation functions defined
in the next section the factors $A(\textbf{p})$ cancel and the formulae are as
if the sources were point-like. Therefore, the correlation function depends
only on $q$ and not on the vectors $p_1,p_2$ separately. This is similar to the
situation in the GGLP model. Note, however, the change in the interpretation of
the dependence of function $\rho(x_s)$ on the space vector $\textbf{x}_s$. Now
it gives the distribution of the positions of the sources. Since the sources
are smeared in space, the actual region, where the pions are produced is larger
than would follow, if the function $\rho(x_s)$ were interpreted as the GGLP
function $\rho(\textbf{x})$ was. What is more, the function of four variables
$\rho(x_s)$ cannot be determined unambiguously from measurements of the
three-momenta. Thus unavoidable model dependence comes in.

Another less important, but much publicized, difference between KP and GGLP was
in the assumption about the space distribution of sources. KP pointed out that
the interaction region is likely to be opaque and that, consequently, only the
sources on its surface can contribute. They assumed that the sources are
created on the surface of a sphere of radius $R$, but that their radiation
satisfies Lambert's law taken to mean that this surface can be replaced by a
circular disc of radius $R$ perpendicular to the direction from the interaction
region to the detector\footnote{This is a hand-waving argument. An exact
evaluation of the corresponding integral can be found in ref. \cite{KP5}. It
yields the result  (\ref{cockop}) in the limit $c\tau \gg R$. This condition
was clearly stated in refs \cite{KOP2}, \cite{KP2}. Unfortunately, it is hardly
ever realized in practice. Another case when the formula is valid is for
$q_\parallel = 0$, where $q_\parallel$ is the component of vector $q$ along the
direction of the total momentum of the pair.}. Thus for the integral in
(\ref{rhoint}) they got

\begin{equation}\label{cockop}
\frac{1}{\pi R^2}\int_0^R dr r \int_0^{2\pi}e^{iq_T\cdot r \cos\theta} =
\frac{2J_1(q_T R)}{q_T R}.
\end{equation}
Here $q_T$ is the length of the component of $\textbf{q}$ in the plane of the disc and
$J_1$ is a Bessel function. The integral is most simply done by expanding the exponent
in the integrand and integrating term by term. Cocconi \cite{COC} proposed to replace
this expression by a Gaussian, according to

\begin{equation}\label{}
  \left(\frac{2J_1(x)}{x}\right)^2 \rightarrow e^{-\frac{x^2}{4}}.
\end{equation}
Actually, for not too large values of the argument, the right hand side is a
very good approximation to the more complicated expression on the left-hand
side. This short reference does not give justice to the importance of reference
\cite{COC}. Cocconi derived independently many of the KP results. He proposed
alternative interpretations. E.g. in his picture $c\tau$ was the thickness of
the "photosphere" i.e. of the spherical crust from which the pions were
emitted. Last not least, he contributed much in discussions with
experimentalists, so that ref. \cite{COC} has got more citations than any of
the KP papers and even sometimes the whole approach was referred to as the
Kopylov, Podgoretskii Cocconi model.

For the distribution in the difference in momenta KP got \cite{KOP2}

\begin{equation}\label{}
\frac{d^3\sigma}{d^2q_Tdq_0} \sim 1 + \left[\frac{2J_1(q_T R)}{q_T
R}\right]^2\frac{1}{q_0^2\tau^2 + 1}.
\end{equation}

Kopylov \cite{KOP2} rewrote this in the form

\begin{equation}\label{kopfor}
\frac{d^3\sigma}{d^2q_Tdq_0} = \left(\frac{d^3\sigma}{d^2q_Tdq_0}\right)_{off}\left( 1
+ \frac{I(q_T R)}{1 + q_0^2\tau^2}\right).
\end{equation}
Note that this formula would not follow, if the second factor on the right-hand
side depended on $\textbf{p}_1 + \textbf{p}_2$, because the two-particle pseudo
density matrix yields $d^6\sigma/dp^6$ and after integration over $\textbf{p}_1
+ \textbf{p}_2$ factorization would not hold any more. Function $I(q_TR)$ can
be obtained by comparison with the previous formula, or from some other model.
The important distinction is between the cross-section on the left-hand side,
which is the physical one containing all the interference effects, and that on
the right-hand-side, which is calculated with the interference effects turned
off. In the GGLP and KP models this would be up to a constant factor
$\rho(\textbf{p}_1,\textbf{p}_1)\rho(\textbf{p}_2,\textbf{p}_2)$, but the
formula is more general than that. Then the "off" cross-section, which was soon
renamed by experimentalists "background", should be taken from experiment. The
question is how? GGLP used for comparison with like-sign pion pairs the
unlike-sign pion pairs. These have certainly no Bose-Einstein correlations, but
they have other unwanted correlations due e.g. to the formation of resonances
in the $\pi^+\pi^-$ system. Kopylov \cite{KOP2} proposed to take like sign
pairs, but with each pion taken from a different event\footnote{According to
ref. \cite{KOP3} this was an idea of Podgoretskii.}. This background got later
the name "mixed" and became quite popular, though it eliminates too many
correlations e.g. the correlations due to energy and momentum conservation or
to the production of jets.

The relation between the work of GGLP and that of Hanbury Brown and Twiss was
clarified in ref. \cite{KP4}. The authors developed a more general model. It
introduces two scales of time, or energy: the life time of the emitting source,
or equivalently the width of the energy distribution of the produced particles,
$\tau \sim \frac{1}{\Gamma}$ and the resolution of the detector in time, or
equivalently in energy, $\Delta t \sim \frac{1}{\gamma}$. In astronomy the life
time of the excited atomic state, which emits the radiation, is much shorter
than the time resolution of the detector. Therefore, one can neglect $\tau$
compared to $\Delta t$. In particle physics, on the other hand, the width in
energy of the source is much larger than the energy resolution of the detector.
Therefore, the opposite limit is justified: one can neglect $\gamma$ as
compared to $\Gamma$. These two limits correspond respectively to the models of
Hanbury Brown and Twiss and to that of GGLP. They are completely different
limits. E.g. the HBT results can be derived from classical physics without ever
mentioning quantum physics. Nevertheless, the acronym GGLP got in the seventies
gradually replaced by the acronym HBT. The analysis from \cite{KP4} implies
that the motivation must have been sociological rather than physical. In this
review we use the neutral BEC, which is also quite popular.

\subsection{Correlation functions}

Since the study of various correlations was a popular subject in
the seventies, it soon became common to express formulae like
(\ref{kopfor}) in terms of two-particle correlation functions, cf.
e.g. \cite{BIS}, defined by

\begin{equation}\label{corfun}
  C(\textbf{p}_1,\textbf{p}_2) =
 {\cal{N}} \frac{N(\textbf{p}_1,\textbf{p}_2)}{N(\textbf{p}_1)N(\textbf{p}_2)}.
\end{equation}
Here

\begin{equation}\label{sindub}
  N(\textbf{p}) = \frac{1}{\sigma}\frac{d^3\sigma}{dp^3};\qquad
  N(\textbf{p}_1,\textbf{p}_2) = \frac{1}{\sigma}\frac{d^6\sigma}{dp_1^3 dp_2^3};
\end{equation}
are respectively the single particle and the two-particle
distributions of the particles studied, $\sigma$ is the integrated
single particle cross-section, and $d^3p$ means either
$dp_xdp_ydp_z$ or $dp_xdp_ydp_z/E_p$. ${\cal N}$ is a
normalization constant. Theorists usually either put it equal one,
or interpret it as the ratio of the normalization of the
denominator to the normalization of the numerator. In the latter
case, for inclusive processes they put ${\cal N} = \frac{\langle n
\rangle^2}{\langle n(n-1)\rangle}$. The resulting confusion has
been reviewed in ref. \cite{MIV}. Experimentalists more often use
this factor to make the correlation function tend to one with
increasing $Q^2$.

The correlation function is a nice, well-defined object, but for extracting
information about the production region it is often "extremely impractical" as
put later by experimentalists \cite{ZAJ}. Consider for example a process, where
the final state consists of two narrow jets back to back. Let the orientation
of the axis of this pencil-like structure have an isotropic distribution. Then
function $N(\textbf{p}_1,\textbf{p}_2)$ has strong maxima for $\theta_{\pi\pi}
= 0$ and $\theta_{\pi\pi} = \pi$, while function $N(\textbf{p})$ is spherically
symmetric. The correlation function also has a strong maximum at
$\theta_{\pi\pi} = 0$ (and another one at $\theta_{\pi\pi} = \pi$), but this
tells us little about Bose-Einstein correlations. On the other hand, if for
each event the coordinate system is rotated so that the axis of the jets is
along the $z$ axis, also function $N(\textbf{p})$ acquires maxima at
$\theta_{\pi\pi} = 0$ and $\theta_{\pi\pi} = \pi$. The maxima in the
denominator of (\ref{corfun}) largely cancel the dynamical part of the maxima
in the numerator and Bose-Einstein correlations can be studied, though usually
some additional selections to improve the analysis are introduced ( c.f. e.g.
\cite{AIH}). Other examples, where the correlation function (\ref{corfun}) is
in practice useless for gaining information about the interaction region, have
been described in ref. \cite{GUL}. Formally, the way out is to define the
correlation function $C_0(p_1,p_2)$ for the background, i.e. for the fictitious
case considered by KP:  the same process, but with the Bose-Einstein
correlations switched off. Then the reduced correlation function

\begin{equation}\label{}
C(\textbf{p}_1,\textbf{p}_2)/C_0(\textbf{p}_1,\textbf{p}_2) \equiv 1 +
R(\textbf{p}_1,\textbf{p}_2).
\end{equation}
is a good replacement for the correlation function (\ref{corfun}). In the
following we will mostly discuss the function $R(\textbf{p}_1,\textbf{p}_2)$
defined by this formula. It corresponds to the interference term in formula
(\ref{kopfor}) and should tend to zero, when $Q^2$ becomes so large that no
more Bose-Einstein correlations are expected. When the background two-particle
distribution is a product of the standard inclusive single particle
distributions, $C_0(\textbf{p}_1,\textbf{p}_2) = 1$ and $1 +
R(\textbf{p}_1,\textbf{p}_2)$ reduces to the correlation function
$C(\textbf{p}_1,\textbf{p}_2)$. This is the case for the mixed background,
while usually it is not the case for the background from unlike sign pion
pairs. It is also not the case when the inclusive single particle distribution
is modified by some selection or transformation like in the two-jet model
mentioned above. Since it is usual to call $1 + R(\textbf{p}_1,\textbf{p}_2)$ a
correlation function, one could argue as follows. The uncorrected experimental
correlation function $C_{uncorr}(\textbf{p}_1,\textbf{p}_2)$ should be
corrected for the unwanted correlations by dividing it by
$C_0(\textbf{p}_1,\textbf{p}_2)$. Then $1 + R(\textbf{p}_1,\textbf{p}_2)$ is
this corrected correlation function. In the seventies theorists usually put
$C_0 = 1$ and experimentalists chose $C_0$ constant and such as to make
$R(\textbf{p}_1,\textbf{p}_2)$ tend to zero with increasing $Q^2$. Later the
sophistication was greatly increased, but the best way of choosing
$C_0(\textbf{p}_1,\textbf{p}_2)$ is even today  controversial.

\subsection{Further experimental results}

On the experimental side some more qualitative implications of the GGLP picture
got confirmed. The effect was found in nucleus-nucleus collisions \cite{FUN}
and for $K^0_sK^0_s$ pairs \cite{COO}. It was shown \cite{ESA} that within the
experimental errors the GGLP effect could be interpreted as a reflection of the
difference in the distributions of the masses of pion pairs $m_{\pi\pi}$ for
pairs of like-sign and unlike-sign pions. Actually GGLP proposed that the
weight factor, which equals one for unlike-sign pairs, should depend on $Q^2 =
-(p_1 - p_2)^2$ for like-sign pairs. Since, however, $Q^2 =  m^2_{\pi\pi} -
4m_\pi^2$, the two statements are equivalent. Three- and four-body correlations
were studied \cite{BMN}, \cite{BOE}, \cite{LES}. They were found strong, but at
that time could be fully explained as reflections of the GGLP two-body
correlations. Gradually the groups begun to determine the radii and life times
of the interaction regions as suggested by KP. Results were obtained for $K^+p$
scattering \cite{DEW}, \cite{GOO}, \cite{GRA}, for $\pi^\pm p$ scattering
\cite{ANG1}, \cite{CAL}, \cite{DEU}, for $\overline{p} p$ \cite{ANGE},
\cite{BEC}, \cite{COO}, \cite{LOK} and $\overline{p}n$ \cite{BOR} scattering ,
for $p p$ scattering \cite{BEC}, \cite{DRI}, \cite{EZE}, for $p$-nucleus ($Be,
Ti, W$) \cite{BEC} and $\pi^--C$ \cite{ANG2} scattering as well as for
nucleus-nucleus scattering \cite{FUN} ($Ar + BaI_2$ and $Ar + Pb_3O_4)$,
\cite{ANG3} (C + Ta). The background distribution was usually identified with
the distribution for unlike sign pairs. Only for scattering on nuclei the mixed
background proposed by Kopylov was found more convenient, because using it one
could use the negatively charged particles only and thus, avoid the
contamination by protons. Sometimes results for two different backgrounds were
compared in order to show that the uncertainty about the background
distribution does not affect too much the results. The normalization of the
background distribution with respect to the distribution of like-sign pairs was
chosen so that the numbers of pairs in both samples coincided, or so that the
distributions became identical for large values of $Q^2$. Note that these two
normalizations are not equivalent. There were various other sources of
uncertainty. Statistics was poor and the results depended on the binning
chosen. The fits, as judged by the $\chi^2$ test, were often poor. The maps of
$\chi^2(R,\tau)$ exhibited large valleys with secondary minima. Formula
(\ref{kopfor}) is not covariant. Some groups used it in the overall cms, others
preferred the cms of the charged pions. On that were superposed the ordinary
experimental uncertainties e.g. particle misidentification and limited momentum
resolution. The theory was not very robust either. E.g. when the authors of ref
\cite{LOK} studying $\overline{p}p$ annihilations got $R \approx c\tau \approx
3$fm and, not very surprisingly, considered this result too large, they
replaced the assumption that $\textbf{v} = 0$ by the assumption that there are
two sources moving with velocities $\textbf{v}$ and $-\textbf{v}$ respectively,
added some phenomenological assumptions about $\textbf{v}$ and got from the
same data $c\tau \approx 0.7$fm and $R \approx 0.7 -  0.8$fm. Another degree of
freedom, which was not exploited by experimentalists however, was to assume
that there is a scatter of the time when the sources got excited. This gives
another factor dependent on $q_0$ \cite{KP2}.

In spite of these uncertainties the results were not unreasonable. Results from
$K^0_sK^0_s$ pairs \cite{COO} were roughly consistent with those from
$\pi^\pm\pi^\pm$ pairs. For hadron-hadron scattering $R$ was usually about one
fermi. For nucleus-nucleus scattering it was somewhat bigger, perhaps $3 - 4$
fm. In both cases $c\tau$ was about one fm, perhaps a bit less. According to
most papers, though not to all, $c\tau < R$ and the interaction region had the
shape of a pancake rather than that of a cigar. Within errors, no systematic
dependence of $R$ and/or $\tau$ on energy and/or the kind of hadrons involved
was seen. The collision energies ranged up to $\sqrt{s} = 52.3$ GeV \cite{DRI},
which would have been impossible to study using the old parameters $\gamma$
which hardly depend on the charges at so high energy per particle \cite{CZS}.

The KP parameterization was not the only one used. In particular Biswas et al. in ref.
\cite{BIS} found

\begin{equation}\label{}
  C(\textbf{p}_1,\textbf{p}_2) = 1 + (0.80 \pm 0.10)e^{-(11.2 \pm 2.4)Q^2}.
\end{equation}
According to this result $R(\textbf{p},\textbf{p}) = 0.8 \pm 0.1$. This was
surprising, as according to both GGLP and KP, as well as to a more general
argument modelled on quantum optics \cite{BAZ}, $R(\textbf{p},\textbf{p}) = 1$.
The experimental evidence for $R(\textbf{p},\textbf{p}) < 1$  was not yet
compelling. Moreover, experimental factors like particle misidentification and
finite momentum resolution could also reduce $R(\textbf{p},\textbf{p})$.
Nevertheless, the problem attracted attention.

\subsection{Further theoretical ideas}

One way to obtain $R(\textbf{p},\textbf{p}) < 1$ is to reject the assumption
that the pion sources associated with space-time points are incoherent. When
all the sources act coherently, they can be replaced by one common source. Then
there is no need to symmetrize and consequently $R(\textbf{p},\textbf{p})=0$ —-
there is no attraction between the momenta of identical bosons. Fowler and
Weiner \cite{FOW}, \cite{FOW1} pointed out that for a partially coherent source
one can obtain $0 < R(\textbf{p},\textbf{p}) < 1$. A surprising conclusion was
that in order to explain the small reduction observed by \cite{BIS} one needs
about $50$\% of coherence. In a more quantitative model proposed in \cite{BIY1}
the coherent sources were responsible for 70\% of the pion production. A
related strategy was to use a more sophisticated model for multiple particle
production, where the symmetry with respect to exchanges of identical pions is
built in from the outset. Work along such lines was done by Giovannini and
Veneziano \cite{GIV}, but unfortunately it gave the prediction
$R(\textbf{p},\textbf{p}) = 0$ for multiple particle production in $e^+e^-$
annihilations, which soon got disproved by experiment.

Implications of the fact that some of the interfering pions were decay products
of resonances have been considered since the beginning of the KP series of
papers \cite{GKP}, \cite{KOP1}. In the mid seventies, however, it was
established experimentally that for multiple particle production in
hadron-hadron scattering the fraction of prompt, i.e produced directly and not
from decaying resonances, pions in multiple particle production processes is
small. Typical estimates were between 10\% and 20\% \cite{GRAS}. This suggested
that interference effects could be used to study resonances, as foreseen
earlier \cite{KOP1}, but also that studies of the Bose-Einstein correlations,
which did not take resonances into account, might be unrealistic \cite{GRAS},
\cite{THO}. In particular Grassberger \cite{GRAS} produced a realistic model
including resonances and worked out many of its predictions. He found that the
GGLP and KP models require important revisions. Since there is a time lag
between the production of the prompt pions and the production of the pions from
resonance decays, the interference between pions of these two origins should
produce narrow peaks which should be added to the more smooth functions
$R(q^2)$ obtained neglecting resonances. This gives
$R(\textbf{p},\textbf{p})>1$, but the peaks corresponding to the long lived
resonances, in particular to the $\eta$ and $\omega$, are so narrow that they
are missed under the experimental conditions and one expects that effectively
in practice $R(\textbf{p},\textbf{p}) < 1$. Another argument in favor of the
narrow peaks is that the resonances fly away before decaying which makes the
region where the pions are created larger. It should be kept in mind, however,
that if the resonances did not move at all, the narrow peaks would still be
there because of the time lag. Since the resonance fractions depend on momenta,
the correlation functions $C(\textbf{p}_1,\textbf{p}_2)$, should depend not
only on the momentum difference $q$, but also on the sum of momenta i.e. on
$K$. The pions from resonance decays, especially from the long-lived
resonances, are produced in a region, where the density of particles is low,
which makes doubtful the claim that pion production happens only on the surface
of the production region. Finally, from the explicit formulae it is seen that
the correlation functions depend on the orientation of the difference of
momenta $\textbf{q}$. For $\textbf{q}$ parallel to the beam axis the peak in
$Q^2$ is narrower than that for $\textbf{q}$ perpendicular to the beam. This
means that in this model the production region is cigar-shaped. A proposal
\cite{GRAS} with a great future was to choose for study, besides the
longitudinal (with respect to the beam) component of $\textbf{q}$, its
transverse component parallel to the transverse momentum of the pion pair,
later called the \textit{out} component, and the transverse component
perpendicular to the transverse momentum of the pion pair, later called the
\textit{side} component. In the examples worked out in the paper $R_{side}$ is
always the smallest. $R_{out}$ for $\textbf{p} = 0$ must be equal $R_{side}$
from symmetry, but with increasing $|\textbf{p}_T|$ it increases and is already
larger than $R_{long}$ at $|\textbf{p}_T| = 300$ MeV/c. $R_{long}$ decreases
significantly when $|\textbf{p}_T|$ increases from $100$ MeV to $300$ MeV. At
the time the experimental data were not good enough to check these predictions.
The differences between the new picture and the GGLP-KP one were so large that
Grassberger concluded: "If they [the peaks] are not found, this would have
serious implications for resonance cross-sections. If they are found, however,
the currently observed $\pi^-\pi^-$ correlations cannot be the particle physics
equivalent of the Hanbury Brown-Twiss effect".

Another influential contribution was the paper by Yano and Koonin \cite{YAK}.
They were interested in heavy ion collisions and their starting point was a
classical single particle distribution

\begin{equation}\label{yankoo}
  D(x,p) = \frac{E}{\sigma}
  \frac{d\sigma}{d^3p}(\pi^2r_0^3\tau)^{-1}\exp\left[-a(x\cdot P)^2 + bx^2\right],
\end{equation}
where $P$ is the total initial four-momentum. The physical origin
of this formula is simple. The measured momentum distribution is
multiplied by a guessed distribution in space-time and the product
is interpreted as the phase-space distribution. Since the source
function $D(x,p)$ depends on both coordinates and momenta it is,
of course, unsuitable for substitution into the Klein-Gordon
equation. The parameters $a$ and $b$ can be rewritten as
\cite{YAK}

\begin{equation}\label{}
  a = s^{-1}\left(r_0^{-2} + \tau^{-2}\right);\qquad b = r_0^{-2}.
\end{equation}
In the overall cms frame $P = (\sqrt{s},\textbf{0})$ and formula (\ref{yankoo})
becomes

\begin{equation}\label{cmsyak}
 D(x,p) =
 \frac{E}{\sigma}\frac{d\sigma}{d^3p}(\pi^2r_0^3\tau)^{-1}
 \exp\left[-\frac{t^2}{\tau^2}-\frac{\textbf{x}^2}{r_0^2}\right].
\end{equation}
The corresponding two-body distribution is

\begin{equation}\label{}
  \frac{E_1E_2}{\sigma}\frac{d\sigma_c}{d^3p_1d^3p_2} = \int d^4x_1 d^4x_2
  D(x_1,\textbf{p}_1)
  D(x_2,\textbf{p}_2)|\phi^s_{\textbf{p}_1,\textbf{p}_2}(x_1,x_2)|^2,
\end{equation}
where $\phi^s_{\textbf{p}_1,\textbf{p}_2}(x_1,x_2)$ is the
symmetrized two-body wave function. In the GGLP and KP models
symmetrized plane waves had been used . Yano and Koonin considered
also Coulomb distorted waves and Coulomb distorted waves including
strong $\pi^-\pi^-$ interactions in the $I =2$, $L = 0,2$ states.
The Coulomb and strong corrections turned out to be marginal
(below 0.1\% in the interesting region of $\textbf{q}$), so
finally only the plane waves were used. These give

\begin{equation}\label{}
  R(\textbf{p}_1,\textbf{p}_2) = \exp\left[-\frac{a}{2} r_0^2\tau^2(q\cdot P)^2 + \frac{1}{2b}q^2\right]
\end{equation}
and in the overall cms system

\begin{equation}\label{yakmom}
  R(\textbf{p}_1,\textbf{p}_2) = \exp\left[-\frac{1}{2}q_0^2\tau^2 -
  \frac{1}{2}\textbf{q}^2r_0^2\right].
\end{equation}
This formula, or formula (\ref{cmsyak}), gives the physical interpretation of
the parameters $\tau$ and $r_0$. As compared to the KP formalism it suggests a
change from the parameter $q_T$ to $\textbf{q}^2$. Note also the change, with
respect to the GGLP formula (\ref{gglcor}), of the sign in the term
proportional to $q_0^2$ in the exponent.

Let us consider now the quantum-mechanical interpretation of the YK formulae. Using
the plane wave approximation for function $\phi^s_{\textbf{p}_1,\textbf{p}_2}$ and the
fact that function $D(x,p)$ is real, we get by comparison with (\ref{rhosym})

\begin{equation}\label{ro12yk}
|\rho(\textbf{p}_1,\textbf{p}_2)|^2 = Re \int d^4x_1D(x_1,p_1)e^{iqx_1}\int
d^4x_2D(x_2,p_2)e^{-iqx_2}.
\end{equation}
This cannot be generally true, since the right hand side does not even have to be
positive. Following YK we assume further, however, that

\begin{equation}\label{quayak}
  D(x,p) = |A(\textbf{p})|^2\rho(x),
\end{equation}
where $\rho(x)$ is real non-negative. Then the $Re$ is unnecessary and one can put

\begin{equation}\label{}
\rho(\textbf{p}_1,\textbf{p}_2) =
A(\textbf{p}_1)A^*(\textbf{p}_2)\left\langle e^{iqx}
\right\rangle,
\end{equation}
which formally coincides with (\ref{rhokop}) and thus, is acceptable as a
formula consistent with quantum mechanics. The factorization of the $D(x,p)$
function assumed in (\ref{quayak}) means that there are no $\textbf{p}-x$
correlations. This is a reasonable assumption \cite{HUM1} for the low energy
heavy ion collisions considered in \cite{YAK}, but not in general.

When there is no factorization, the YK integrands contain the non-factorizable
analogues of $\rho(x_1)|A(\textbf{p}_1)|^2$ and $\rho(x_2)|A(\textbf{p}_2)|^2$
respectively, while in quantum mechanics the integrands contain the non
fatorizable analogue of $\rho(x)A(\textbf{p}_1)A^*(\textbf{p}_2)$ and its
complex conjugate. This can give similar results only under very special
assumptions (cf. \cite{PGG}). In order to see how the YK approach can break
down when there are momentum-position correlations, let us consider the
following example\footnote{This is a simplified version of the models described
in refs \cite{PRW2}, \cite{MKF}. See also \cite{CSZ}.}. Replace in formula
(\ref{yankoo}) the normalized Gaussian by $\delta^4(x - \lambda p)$, where
$\lambda$ is a constant. Then the integrals in formula (\ref{ro12yk}) can be
done and one finds

\begin{equation}\label{}
  |\rho(\textbf{p}_1,\textbf{p}_2)|^2 =
  \rho(\textbf{p}_1,\textbf{p}_1)\rho(\textbf{p}_2,\textbf{p}_2)
  \cos\left(\lambda
  Q^2\right),
\end{equation}
which is evidently wrong since the right-hand side is not positive
definite. For some more realistic models where the YK method fails
because of the $x$ -- $\textbf{p}$ correlations and for its
derivation using the "smoothness assumption" see \cite{PRA6}.

\subsection{The GKW paper}

A detailed discussion of particle correlations in heavy ion collisions was given by
Gyulassy, Kauffmann and Wilson (further quoted GKW) \cite{GKW}. This is the second
most quoted paper (after GGLP \cite{GGL}) on Bose-Einstein correlations in multiple
particle production processes. The starting point of GKW was the Klein-Gordon equation
with a classical current on the right-hand side (\ref{klegor}), but with $A$
interpreted as a Heisenberg pion field. The normalized solution, which had been well
known, is the coherent state

\begin{equation}\label{solgkw}
  |A\rangle = e^{-\overline{n}/2}\exp\left(i\int d^3p\tilde{J}(\textbf{p})
  a^\dag(\textbf{p})\right)|0\rangle,
\end{equation}
where $a^\dag(\textbf{p})$ is the creation operator for a pion with momentum
$\textbf{p}$,

\begin{equation}\label{}
  \tilde{J}(\textbf{p}) = \int d^4x \frac{e^{iE_pt - i\textbf{p}\cdot \textbf{x}}}
  {(2E_p(2\pi)^3)^{1/2}} J(x),
\end{equation}
 and

\begin{equation}\label{}
  \overline{n} = \int d^3p |\tilde{J}(\textbf{p})|^2.
\end{equation}
The use of coherent states to describe Bose-Einstein correlations is standard in
optics and had been applied to multiple hadron production processes \cite{BAZ}, but
GKW pushed the analysis much further than their predecessors.

Let us denote by $|J\rangle$ the coherent state corresponding to the current $J(x)$.
Since as is well known

\begin{equation}\label{}
a^\dag(\textbf{p})|J\rangle  = i\tilde{J}(\textbf{p})|J\rangle,
\end{equation}
one easily checks that \cite{GKW} in state $|J\rangle$ the multiplicity distribution
is Poissonian with average multiplicity $\overline{n}$. For a given multiplicity $m$
the momentum distribution is

\begin{equation}\label{}
N(\textbf{p}_1,\ldots,\textbf{p}_m) = \prod_{j=1}^m |\tilde{J}(\textbf{p}_j)|^2
\end{equation}
and consequently $R(\textbf{p},\textbf{p}) = 0$ - there is no GGLP effect. Now the
authors introduce $N$ source currents related by space-time translations. The density
matrix is assumed to be diagonal in $N$ and the probability distribution $P(N)$ is
part of the theoretical input.  Thus the overall source current for given $N$ is

\begin{equation}\label{}
  J(x) = \frac{1}{\sqrt{N}}\sum_{j=1}^N J_\pi(x - x_j),
\end{equation}
or equivalently

\begin{equation}\label{}
  \tilde{J}(\textbf{p}) = \tilde{J}_\pi(\textbf{p})\frac{1}{\sqrt{N}}\sum_{j=1}^N
  e^{i\omega_k t - i\textbf{p} \cdot \textbf{x}},
\end{equation}
where $\tilde{J}_\pi(\textbf{\textbf{p}})$ is the Fourier transform of $J_\pi(x)$. The
factor $1/\sqrt{N}$ was not used by GKW. It makes, however, the limits $N \rightarrow
\infty$ easier to see (cf. e.g. \cite{BAZ}). The probability distribution in
space-time for the points $x_j$ is given by some function $\rho(x)$. It is convenient
to define its Fourier transform

\begin{equation}\label{}
  \rho(p) \equiv \rho^*(-p) =  \int d^4x \rho(x)e^{ipx}.
\end{equation}
Note the crucial difference with respect to the KP approach: it is \textit{not
assumed} that the constituent sources $J(x - x_j)$ are incoherent. Thus the solution
(\ref{solgkw}) can be used with the new current.

Now the single-particle momentum distribution is

\begin{equation}\label{}
  N(p) = |\tilde{J}_\pi(p)|^2\left( 1 + \langle \frac{N+1}{N} \rangle |\rho(p)|^2
  \right),
\end{equation}
where $\langle \cdots \rangle$ means averaging over the
probability distribution $P(N)$. One of the important points of
the GKW paper is the observation that since realistic functions
$\rho(p)$ decrease rapidly with the increase of any of the
components of $p$, they used as a guide formula (\ref{cmsyak}),
and since $E_p \geq m_\pi$ the second term on the right hand side
can be neglected. Another argument for the same conclusion is that
function $\rho(p)$ is small, when $|\textbf{p}|$ exceeds $1/R$,
where $R$ is the radius of the region where the pions are
produced, while typical values of $|\textbf{p}|$ are of the order
of $m_\pi$ or more. Therefore, in the large source limit, which is
a reasonable picture of heavy ion collisions, the second term is
negligible. It is a correction term which is calculable, can and
should be studied, but as a reasonable approximation one can omit
it. Then the result is that on the average $N$ sources contribute
just $N$ times as much as one source, as if the sources were
incoherent. In the two-body distribution there are six kinds of
terms, but after simplifying as in the previous case only two
kinds survive: those which do not contain $\rho(\ldots)$ and those
which contain $\rho(q)$, where all the components of the argument
can be made small. One gets

\begin{equation}\label{}
  R(p_1,p_2) = \langle \frac{N-1}{N}\rangle |\rho(q)|^2.
\end{equation}
For large values of $\langle N \rangle$ the first factor on the right-hand side
is approximately equal one and the GGLP result is recovered. Since similar
results hold for any $m$-pion distribution, one can assume, as a mathematical
shortcut, that the constituent sources are incoherent. GKW called this
approximation the chaotic field limit and pointed out that this limit, though
not the finite $N$ corrections to it, had been found from essentially the same
model in ref. \cite{BAZ}. Because of their careful handling of $N$, however,
GKW were able to derive the incoherence of the sources as a reasonable
approximation, while in ref. \cite{BAZ} it was introduced as an assumption. GKW
realized that the interaction region depends on the impact parameter of the
collision, which they consider to be well-defined in a collision. Thus, they
recommended either to use data corresponding to a narrow range of impact
parameters, or to average incoherently over the impact parameters.

GKW estimated that resonance production is not very important in
heavy ion collisions. Therefore, in order to explain the fact that
$R(\textbf{p},\textbf{p}) \neq 1$ they introduced partial
coherence \cite{FOW}. Generalizing the proposal of Fowler and
Weiner they introduced the degree of coherence for momentum
$\textbf{p}$ by the formula

\begin{equation}\label{}
  D(\textbf{p}) = \frac{n_0(\textbf{p})}{n_0(\textbf{p}) + n_{ch}(\textbf{p})},
\end{equation}
where $n_0$ refers to the pions produced coherently and $n _{ch}$ to the pions
produced incoherently (chaotically). This function should be obtained from the
relation

\begin{equation}\label{}
  R(\textbf{p},\textbf{p}) = 1 - D^2(\textbf{p}).
\end{equation}
The degree of coherence must be known before conclusions about the geometry of
the source can be extracted from the data, because\footnote{Assuming that
$\rho(p)$, $\tilde{J}_0(p)$ and $\tilde{J}_{ch}(p)$ are real.} e.g.

\begin{eqnarray}\label{}
  C(\textbf{p}_1,\textbf{p}_2) & = & 1 + (1-D(\textbf{p}_1))(1-D(\textbf{p}_2))
  \rho^2(q)\nonumber \\
  &&+2\sqrt{D(\textbf{p}_1)D(\textbf{p}_2)(1-D(\textbf{p}_1))(1-D(\textbf{p}_2))}\rho(q).
\end{eqnarray}
Only for $D \equiv 0$ one recovers the result (\ref{rhosym}).   The origin of the
coherent component was ascribed to some collective effects in the colliding nuclei.
Thus it would be specific to heavy ion collisions.

The most influential part of the GKW paper was their analysis of the final
state interactions, i.e. of the interactions taking place after the pions had
been produced. These interactions can affect the pion momentum distribution.
Final state interactions were considered before \cite{GRAS}, \cite{YAK}, but
always with the conclusion that they are unimportant. GKW proposed a model for
these interactions and when it later became used by experimentalists the
corrections were often found significant. There are two kinds of final state
interactions. The pions move in a single particle potential including the
Coulomb potential from the positive charge of the nuclear cores. Moreover, the
pions interact with each other. Here one can consider the Coulomb interactions
between the charged pions and their strong interactions \cite{YAK}. Taking into
account the single particle potential, say $V(x)$, is comparatively easy,
because it does not affect the independent particle approximation. One assumes
that the potential $V(x)$ does not produce particle pairs and does not support
bound states. Then it is enough to replace everywhere the pion plane waves,
which are solutions to the free particle Klein-Gordon equation, by the
corresponding solutions of the Klein-Gordon equation with the potential $V(x)$.
We will denote these solutions $\psi_\textbf{p}(x)$, where the subscript
$\textbf{p}$ means that this function corresponds to the plane wave describing
a free pion with momentum $\textbf{p}$, i.e. that for $\textbf{x}^2 \rightarrow
\infty$ it tends to this plane wave. With the additional assumption that the
range of the single particle potential $V$ is much larger than $m_\pi^{-1}$,
one finds the single particle density matrix

\begin{equation}\label{}
\rho_V(\textbf{p},\textbf{p}') = \int d^4x \rho(x)\psi_p(x)\psi^*_{p'}(x).
\end{equation}
From this density matrix the correlation function is built as
usual, using formula (\ref{rhosym}) to construct the necessary
elements of the two-body pseudo density matrix. The density matrix
may look similar to the density matrix from the KP model with
extended sources (\ref{kproex}), but the physical interpretation
is very different. In the KP model the functions $\langle
x|\psi(x_0)\rangle$ can be normalized to one and describe the
shape of a source with a finite space extension. Here the
functions $\psi_\textbf{p}(x)$ are almost plane waves extending
over all space. They are only slightly deformed in a finite region
of space due to the potential $V(x)$. The effect of coherence is
calculated as before, but using the distorted functions
$\rho_V(\textbf{p},\textbf{p}')$, $n_{0V}(\textbf{p})$ and
$n_{chV}(\textbf{p})$. The distorted functions are calculated with
the plane waves replaced by the functions $\psi_\textbf{p}(x)$. In
practice this part of the final state interaction does not affect
much the distribution of the momentum difference $\textbf{q}$,
because both bosons are pushed in the same way by the potential.

The two-pion interactions spoil the independent particle
approximation and, therefore, are much more difficult to include.
GKW considered various approximations. They pointed out that the
distorted wave Born approximation would be the best tool, but it
is rather complicated. The main effect is to change the wave
function of the pion pair from a symmetrized product of plane
waves to a two-body Coulomb wave function. Using the
nonrelativistic Schr\"odinger equation, factoring out the motion
of the pair as a whole and neglecting in the relative coordinate
space the changes of the wave function in a region of radius
$O(R)$ where $R$ is the radius of the interaction region around
the point $\textbf{x}_1 - \textbf{x}_2 = 0$, one finds \cite{GKW}
that the pseudo density matrix element
$\tilde{\rho}(\textbf{p}_1,\textbf{p}_2;
\textbf{p}_1,\textbf{p}_2)$ gets multiplied by the "well-known
Gamov factor"

\begin{equation}\label{gamov}
  G(\textbf{p}_1,\textbf{p}_2) = \frac{2\pi\eta}{e^{2\pi\eta} - 1}; \qquad \eta = \frac{\alpha
  m_\pi}{|\textbf{p}_1 - \textbf{p}_2|},
\end{equation}
where $\alpha$ is the fine structure constant. According to the
theory the measured two-particle distribution contains this
factor. Thus, the suggestion is that experimentalists could
correct their results for Coulomb interactions by dividing the
measured functions
$\rho(\textbf{p}_1,\textbf{p}_2;\textbf{p}_1,\textbf{p}_2)$ by the
corresponding Gamov factor.

Let us look at the numbers. The Bohr radius for a pair of
identical particles of mass m and unit charge is

\begin{equation}\label{}
  a_B = \frac{2}{\alpha m},
\end{equation}
where $\alpha = 1/137$. This is $1.96$ MeV$^{-1}$ or $386$ fm for
charged pions and $0.56$ MeV$^{-1}$ or $110$ fm for charged kaons.
Since the Bohr radii are much larger than the interactions radii
$R$, which do not exceed several fermis, it had long been believed
that the Coulomb final state interactions can be safely neglected.
From the formula for the Gamov factor it is seen, however, that
the dimensionless parameter is not $R/a_B$, but $1/(qa_B)$, which
can be large when $q$ is small. Moreover

\begin{equation}\label{}
  2\pi\eta = \frac{4\pi}{qa_B} \equiv \frac{c}{q}.
\end{equation}
The coefficient $c$ is $25$ MeV for pions and $7$ MeV for kaons.
It is enhanced by the $4\pi$ factor.

\section{Refinements and criticism in the eighties}

\subsection{New experimental results and problems}

In the eighties the Bose-Einstein correlations were studied in
further processes. They were found in $\mu p$ \cite{ARN} and $\nu
D$ \cite{ALL} collisions and studied in detail in $e^+e^-$
annihilations  \cite{AIH}, \cite{ALT1}, \cite{ALT2}, \cite{AVE},
\cite{JUR} and $\gamma \gamma$ scattering \cite{JUR}. The radii of
the hadron production regions in these processes, as compared to
hadronic interactions, were found somewhat smaller, as expected.
The values of $R(p,p)$, however, were not reduced. This
contradicted the prediction from some models (cf. e.g. \cite{GIV})
that $e^+e^-$ annihilations should be much more coherent than
hadronic interactions and consequently exhibit little or no GGLP
effect. In fact, it made the interpretation of $\lambda$ as the
coherence parameter doubtful.  Correlations between charged kaons
have been found and studied \cite{AKE2}. In general, there was a
spectacular improvement in the quality of the data.

As the experimental data improved the uncertainties and
ambiguities in the analysis of Bose-Einstein correlations became
apparent. In ref. \cite{DEU1} the important result \cite{BIS} that
$R(p,p) \neq 1$ was confirmed. The authors used a formula similar
to the KP formula (\ref{kopfor})

\begin{equation}\label{kpdform}
\frac{N_L}{N_{bcg}} = \mbox{Const}\left(1 +
\lambda\frac{I^2(q_TR)}{1 + (q_0\tau)^2}\right),
\end{equation}
where $I(x) = 2J_1(x)/x$. The introduction of the factor $\lambda$ into the
interference term significantly improves the fits to experimental data and was
soon followed by most groups. Note that in such parameterizations
$R(\textbf{p},\textbf{p}) = \lambda$. Where we write in the following
$R(\textbf{p},\textbf{p})$, many writers would have written $\lambda$. The
interpretation of this coefficient, however, remained doubtful. Experimental
uncertainties, coherence and resonance production were all invoked. The
left-hand side is here the ratio of the distribution for pairs of like sign
pions to the corresponding background distribution of pion pairs. The strongest
result in ref. \cite{DEU1} was that for pairs of like sign pions produced in
$\pi^+p$ interactions at $16$ GeV/c incident momentum: $\lambda = 0.49 \pm
0.02$. Statistically this is about 25 standard deviations from one, but
changing the background from pairs of unlike charge pions to another
background, which looked at least as reasonable though later it was criticized
\cite{ADA},  the authors obtained $\lambda = 0.88 \pm 0.02$. Introducing the
Gamov factor, which was not done in ref. \cite{DEU1}, further significantly
increases the fitted values of the parameter $\lambda$ \cite{BEA1},
\cite{CHA2}, \cite{ZAJ}. Incidentally, introducing the Gamov factor makes the
fits worse. In a particularly bad case \cite{ZAJ} the confidence level dropped
from $65$\% to $0.1$\%. This explains why not all the groups were willing to
include this correction factor. The uncertainty in $\lambda$ was particularly
striking for $\overline{p}p$ annihilations at rest \cite{DEU1}. With the
background of unlike sign pion pairs, $\lambda = 1.20 \pm 0.08$ was the biggest
among the three $\lambda$-s for the reactions considered in the paper. With the
other background it became $0.63 \pm 0.05$ - the smallest of the three.  For
$Q^2$ decreasing towards zero function $R(p_1,p_2)$ was found to increase
faster than the Gaussian fit suggested. This was described either, following
\cite{LEP}, by including in the fit another Gaussian with a larger value of $R$
 \cite{AKE4}, \cite{AKH1},  \cite{ANG4}, \cite{BAI}, \cite{KUL}, or by using
exponentials in $Q$ instead of Gaussians \cite{AIH}, \cite{KUL}, \cite{LOR}.
Taking into account this additional rise again increased the value of
$\lambda$. The uncertainty in $\lambda$ is seen to be large and mainly
systematic. It was not clear how to reduce it or even how to estimate it. This
has made progress in the understanding of the origin of this factor difficult.
The key problem is a good choice of the background distribution.

Let us note some new ideas concerning the evaluation of the background
distributions. In order to eliminate the effects of resonances and their
reflections, the measured distributions for like meson pairs and those for the
background were divided by the corresponding Monte-Carlo distributions
\cite{ALT1}, \cite{ALT2}, \cite{ARN}. A factor $(c_0 + c_2 Q^2)$  \cite{ADA},
\cite{ALT1}, \cite{ARN}, \cite{BRE}, \cite{JUR}, $(c_0+ c_1 Q)$ \cite{AIH}
(with constant $c_i$-s), or similar  \cite{ALT2}, \cite{BAI}, were introduced
into $C_0(p_1,p_2)$ to keep the correlation function close to one in the
region, where Bose-Einstein correlations are no more expected. The groups using
mixed backgrounds realized that, in order to get a good mixed background,
members of the pairs of events supplying the background pairs must be similar.
Depending on the process this could mean similar multiplicity (cf. e.g.
\cite{ADA}, \cite{BEA1}, \cite{BEA2}), rotating one of the events to make the
sphericity axes coincide (cf. e.g. \cite{AIH}) etc. The mixed distribution was
sometimes  \cite{AIH}, \cite{ZAJ} improved as follows. From (\ref{corfun}) one
finds

\begin{equation}\label{}
  N(p_1) = {\cal N}' \int d^3p_2 \frac{N(p_1,p_2)}{1+R(p_1,p_2)},
\end{equation}
where ${\cal N}'$ is a suitable normalizing factor. In first
approximation $R(\textbf{p}_1,\textbf{p}_2) = 0$ and this is just
the single particle distribution as used to build the ordinary
mixed background. One can do better, however, by substituting into
the integrand the function $R(\textbf{p}_1,\textbf{p}_2)$
calculated in the first approximation. After a few iterations one
gets a single particle distribution which satisfies the equation
above. This, however, is a small correction. Note that the product
of such single particle distributions still does not contain the
two-body non Bose-Einstein correlations between identical
particles, which should be there.

Since the results for the life-time $\tau$ were also usually found with large
uncertainties, most of the discussion concerned the radii $R$ of the interaction
regions. As a crude characteristics one could say that within the large uncertainties
all these radii are similar except for the much bigger radii found in heavy ion
collisions. Nevertheless, it was possible to find some interesting regularities.
Indications were found \cite{AKE2}, \cite{COO} that when identical kaons are used, the
radii come out smaller than the corresponding radii obtained when identical pions are
used. This could mean a decrease of the effective (measured) radius with increasing
mass of the identical particles.

The increase of $R$ with increasing particle multiplicity or particle density
in pseudorapidity was at first somewhat confusing. For hadron-hadron
collisions, when increasing the particle multiplicity by selection or by rising
the energy, an increase of $R$ was visible, but only at energies $\sqrt{s} >
30GeV$ \cite{AKE1}, \cite{AKE2}, \cite{AKE3}, \cite{ALB}, \cite{BRE}. For
$e^+e^-$ collisions at energies $\sqrt{s} < 30$ GeV no increase was seen.  For
heavy ion collisions increasing the mass number $A$, or reducing the collision
parameter gave an increase of $R$ even for energies of the order of $1$ GeV
\cite{AGA}, \cite{AKH2},\cite{BEA3}, \cite{LUJ}. What counts here is the mass
number of the projectile, provided the target is heavy enough \cite{AGA}. This
has been nicely illustrated in ref. \cite{DEM}, where for $pp$ and $p+Xe$
scattering the radii are the same within errors, while the particle density in
rapidity changes by a factor of two. Refs. \cite{BAM}, \cite{HUM2} reported for
${}^{16}O + Au$ collisions a strong increase of $R_T$ (alomst by a factor of
two) for rapidities of the pion pairs close to the cms rapidity, but this was
not confirmed. From a thorough compilation of data for heavy ion collisions it
was concluded \cite{BAR1} that the measured radius $R$ increases linearly with
the cubic root of the mass number of the lighter of the two colliding nuclei. A
fit gave $R = 1.21 A_p^{\frac{1}{3}}$. Some evidence for the decrease of the
measured radius with increasing momentum \cite{ANG4} and transverse momentum
\cite{BEA3} of the pion pair was also reported.

Another interesting observation was that for $e^+e^-$ \cite{ALT2}, \cite{AVE},
\cite{JUR} and hadron-hadron \cite{AKE3}, \cite{AKE4}, \cite{BRE} interactions,
$R(p_1,p_2)$ could be well approximated by a function of $Q^2$ only -- in first
approximation by a Gaussian as proposed by GGLP. This has in particular two
important implications. Contrary to the Ansatz made for heavy ion collisions in
ref. \cite{YAK}, for $e^+e^-$ annihilations the enhancement due to
Bose-Einstein correlations is large for $Q^2$ small even if $q_0^2$ and
$\textbf{q}^2$ taken separately are large. The other conclusion is that the
distribution of $\textbf{q}$ in the rest frame of the pair is spherically
symmetric. In the KP formula the increase of either $q_0^2$ or $q_T^2$ reduces
$R(p_1,p_2)$ like in \cite{YAK}, but because of the absence of the dependence
on $q_\|$ (of the component of $\textbf{q}$ along the direction of
$\textbf{p}_1 + \textbf{p}_2$) the fits are much better than in the YK case.
The reason is that \cite{ALT2},\cite{JUR}

\begin{equation}\label{}
  e^{-\frac{R^2}{4}q_\parallel^2 + \frac{\gamma^2R^2}{4(\gamma^2+1)}q_0^2} \approx 1,
\end{equation}
where $\gamma$ is the Lorentz factor for the transformation from
the cms to the rest frame of the pion pair. Introducing this
factor into the KP formula for the correlation function is
approximately equivalent to a change from $\textbf{q}_T^2$ to
$\textbf{q}^2$ accompanied by the introduction of a term, which
increases with increasing $q_0$. Thus the KP formula can be
rewritten so that it becomes similar to the GGLP formula.

\subsection{String model}

For $e^+e^-$ annihilations, where the final state is believed to evolve from a
single string, the effects of Bose-Einstein correlations were calculated
\cite{BOW1}, \cite{ANH} in the framework of a slightly extended \cite{ANH},
\cite{ARB} string model. The string model is closely related to the realistic
and highly successful LUND model \cite{AND}. Therefore, its predictions
concerning the distribution and properties of particle sources must be taken
seriously. Already the qualitative results were striking. There are strong
correlations between the position of the source and the momentum spectrum of
the particles it emits. As a result, the volume determined from the study of
Bose-Einstein correlations is not the volume of the interaction region. Only
particles with similar momenta contribute to the GGLP effect. Therefore, the
measured volume is the volume of the region, where pions with similar momenta
are being produced and not the total interaction volume. This has little effect
on the transverse dimensions, but the measured longitudinal dimension becomes
almost energy independent, while the true length of the interaction region is
believed to grow roughly like $\sqrt{s}$. In order to introduce
$\textbf{x}-\textbf{p}$ correlations into the KP sources \cite{KP3} it is
enough to remove the assumption that the sources are equivalent \cite{BOW1},
but to find a quantitative relation between the string model and the KP model
generalized in this way is not easy. The attempt in ref. \cite{BOW4} does not
look very convincing. Thus the geometrical interpretation of the measured
volume is somewhat doubtful.  The one-dimensional string model, which is more
constrained than the three-dimensional version, suggests that the space-time
distribution of sources should depend mainly on the variable $\tau_1^2 = t^2 -
z^2$. A natural extension to three dimension is to assume that this
distribution depends on $\tau^2 = t^2 - \textbf{r}^2$, which according to the
standard method of analysis corresponds to a correlation function depending
only on the variable $Q^2$ \cite{BOW1}. This is supported by experiment. The
model suggests also that the distribution in $Q$ is more peaked at small $Q$
and has a longer tail at large $Q$ than a Gaussian. At the quantitative level
the comparison of the model with experiment is difficult and requires the use
of Monte Carlo programs and additional assumptions e.g. about resonance
production. Nevertheless, good agreement with the data from ref. \cite{AIH} was
obtained \cite{ANH}, \cite{BOW3}, though after an interesting detour. At first
it seemed \cite{ANH} that the introduction of long lived resonances ruins the
agreement. Then it was pointed out \cite{BOW3} that the amount of $\eta'$ (0.41
per event) commonly assumed at that time was only a guess and probably a very
bad one. Neglecting $\eta'$ production good agreement with experiment was
obtained. Later experiment confirmed that the production of $\eta'$ in $e^+e^-$
annihilations is indeed small.

\subsection{Wigner functions, emission functions and covariant currents }

It has always been a problem how to put together the available information
about the space-time structure of the interaction region and about the momentum
distribution of the produced particles. Usually in quantum mechanics one works
in the momentum representation or in the coordinate representation, but not in
both simultaneously. KP suggested to consider the space-time distribution of
the points labelling the incoherent sources and a wave function in the momentum
representation, the same for each source \cite{KP3}. Yano and Koonin \cite{YAK}
introduced a distribution $D(x,p)$ without bothering about its interpretation
in quantum mechanics. Pratt \cite{PRA1} proposed to use for the sources the
Wigner function  (cf. \cite{HCS} and references quoted there). The Wigner
function $W(X_1,\textbf{K}_1)$ can be defined by either of the two equivalent
formulae:

\begin{eqnarray}\label{surfun}
  W(X_1,\textbf{K}_1) & = &
  \int \frac{d^3q_1}{(2\pi)^3}\rho(\textbf{K}_1+\textbf{q}_1/2,\textbf{K}_1-\textbf{q}_1/2,t_1)
  e^{i\textbf{q}_1\cdot\textbf{X}_1} \\
  W(X_1,\textbf{K}_1) & = &
  \int \frac{d^3Y_1}{(2\pi)^3}\rho(\textbf{X}_1+\textbf{Y}_1/2,\textbf{X}_1-\textbf{Y}_1/2,t_1)
  e^{-i\textbf{Y}_1\cdot\textbf{K}_1}.
\end{eqnarray}
 Inverting the first equation one finds

\begin{equation}\label{}
  \rho(\textbf{p}_1,\textbf{p}_1',t_1) = \int d^3X e^{-i\textbf{q}_1\cdot
  \textbf{X}_1}W(X_1,\textbf{K}_1).
\end{equation}
After freeze-out, the time evolution is by definition that for
free particles. Therefore, it is convenient to factor out from the
density matrix the free particle evolution factor \cite{PRA2}, or
equivalently to go into the interaction picture. This yields

\begin{equation}\label{}
\rho_{I}(\textbf{p}_1,\textbf{p}_1',t) = \int d^3X
e^{iq_1X_1}W_I(\textbf{X}_1,\textbf{K}_1),
\end{equation}
where in the exponent stands the Lorentz invariant product of
four-vectors. Further the subscripts $I$ will be skipped.

Now come the characteristic assumptions of this approach: the sources are
labelled by their freeze out time $t_0$, and the summation over the sources,
assumed incoherent, is reduced to an integration, with a suitable weight
factor, over the freeze out time $t_0$. Since for any source
$\rho_{I}(\textbf{p},\textbf{p}',t)$  does not depend on the time $t$ for $t
\geq t_0$, where $t_0$ is the freeze out time of this source, one can put $t =
t_0$ on the right hand side. Then the formula for the full single particle
density matrix at any time $t$ larger than the latest freeze out time, reads

\begin{equation}\label{emifun}
 \rho(\textbf{p}_1,\textbf{p}_1') = \int d^4X
e^{iq_1X}S(X,\textbf{K}_1),
\end{equation}
where the matrix element on the left hand side is time independent and the time
component $X_0$ on the right hand side is understood as the freeze-out time
$t_0$ of the source labelled $t_0$.  As seen from this derivation function $S$,
which will be further referred to as the emission function\footnote{Also known
as Wigner function, pseudo Wigner function, source function etc. The name
emission function was used in ref. \cite{PRA2}. }, can be understood as a
product of the Wigner function for the particles produced by the source $t_0$
and of the density factor necessary to convert the summation over the sources
into the integration over $t_0$. The point is, however, that besides this
emission function there is an infinity of other emission functions which, when
substituted into this formula, yield the same density matrix and therefore are
equally acceptable. In particular a small source with a large life-time can
give the same correlation function as a large source with a short life-time
\cite{PRA2}. The statement \cite{PRA1} that $S(x,\textbf{p})$ "can be
identified as the probability of emitting a pion of momentum $\textbf{p}$ from
space-time point x" suggests a way of thinking about the source function which
has become very popular. What can be proved, however, is much less: the Wigner
function $W(x,\textbf{p})$ when integrated over space ($d^3x$) gives the
momentum distribution at time $t$ and when integrated over momenta ($d^3p$)
gives the distribution in space-time \cite{HCS}. Nevertheless, it soon became
customary to discuss the source function using Pratt's picture.

When sources freezing out at different times contribute coherently
the relation of the emission function to Wigner functions becomes
rather tenuous \cite{ZAL3}. In this case it seems better to relate
the emission function to the history of the sources rather than to
some states of the system \cite{PCZ},\cite{CHH}. Let us define the
emission function in space-time by the formula (c.f. formula
(\ref{shurya}))

\begin{equation}\label{}
  \tilde{S}(X_1,Y_1) = \sum_i\left\langle
  J_i(x_1)J_i^*(x'_1)\right\rangle.
\end{equation}
Then according to Shuryak's formula (\ref{shkqxy})

\begin{equation}\label{sxyrho}
  \tilde{\rho}(\textbf{p}_1,\textbf{p}'_1) = \int
  \tilde{S}(X,Y)e^{iK_1Y+iq_1X}d^4Xd^4Y,
\end{equation}
Formula (\ref{emifun}) is reproduced for

\begin{equation}\label{emipcz}
S(X_1,K_1)= \int\tilde{S}(X_1,Y)e^{iK_1Y}d^4Y.
\end{equation}
This is an unambiguous definition of the emission function. The emission
function is real, because complex conjugation is equivalent to a change of the
sign of $Y$ and this can be compensated by a change of the integration
variables from $Y$ to $-Y$. Note, however, that in order to calculate this
emission function it is not enough to know the state of the system at some
time,  or even to know the state of the system at every time. Phase relations
among the states generated at different times are also important. The quantity
being averaged depends on two times, and not on one as would be case for state
vectors, density matrices or Wigner functions.

Using the formula for the single particle density matrix and
formula (\ref{rhosym}) one finds

\begin{equation}\label{}
  C(\textbf{p}_1,\textbf{p}_2) = 1 +
  \frac{\left|\int d^4X S(X,\textbf{K})e^{iqx}\right|^2}
  {\int d^4X S(X,\textbf{p}_1)\int d^4X' S(X',\textbf{p}_2)}.
\end{equation}
Often it is considered an acceptable approximation to replace each
of the momenta in the denominator by $\textbf{K}$. Then the
formula takes a simpler form

\begin{equation}\label{pracor}
C(p_1,p_2) = 1 + \left|\frac{\int d^4x S(x,\textbf{K})e^{iqx}}{\int d^4x
  S(x,\textbf{K})}\right|^2.
\end{equation}
The second term on the right hand side is the function
$R(\textbf{p}_1,\textbf{p}_2)$.

An alternative approach has been worked out by Gyulassy and
collaborators \cite{GKW}, \cite{KOG}, \cite{GYP}, \cite{PAG},
\cite{PGG}. It is an extension of the labelled wave packets method
of KP \cite{KP3}. Suppose that each source (wave packet) is
labelled by a space-time point $x_s$ and a momentum
$\textbf{p}_s$. The energy corresponding to this momentum is
obtained from the condition $p_s^2 = m^2$. Then one assumes that

\begin{equation}\label{}
  \rho(\textbf{p}_1,\textbf{p}_2) = \left\langle
  e^{iqx_s}\rho_{p_s}(p_1,p_2)\right\rangle,
\end{equation}
where the averaging is a summation over the sources $s$. The new point, as compared to
the KP formula (\ref{rhokop}), is that the single source density matrix on the
right-hand side depends on $p_s$. A particularly important special case is when

\begin{equation}\label{covcur}
\rho_{p_s}(\textbf{p}_1,\textbf{p}_2) =
j(\frac{p_sp_1}{m})j^*(\frac{p_sp_2}{m}).
\end{equation}
Interpreting $p_s$ as the average momentum of the pions emitted by the source
this means that each source looks the same in its rest frame, while for KP each
source looked the same in the overall frame of the event (except for a
space-time translation). Also the condition $p_s^2 = m^2$ means that the labels
correspond to the emitted pions rather than to a heavy source. Further, it is
assumed that the averaging over the sources can be expressed as an integration
with weight function $D(x_s,p_s)$. Thus

\begin{eqnarray}\label{}
\tilde{\rho}(\textbf{p}_1,\textbf{p}_2) = \int d^4x_s \int d^3p_s
D(x_s,p_s)\rho_{p_s}(\textbf{p}_1,\textbf{p}_2) = \nonumber \\\int
d^4x_s \int d^3p_s
D(x_s,p_s)j(\frac{p_sp_1}{m})j^*(\frac{p_sp_2}{m})
\end{eqnarray}
This version of the model is known as the covariant current
formalism. A more general model, where the single source density
matrix does not factorize, i.e. does not correspond to a pure
state, and the second equality does not hold, was considered in
\cite{PGG}. The authors proposed

\begin{equation}\label{}
\rho_{p_s}(\textbf{p}_1,\textbf{p}_2) =  N \exp\left[\frac{\left(p_s -
(p_1+p_2)/2\right)^2}{2\Delta p^2} + \frac{(p_1-p_2)^2\Delta x^2}{2}\right],
\end{equation}
where $N$ is the normalization constant, while $\Delta x^2$ and $\Delta p^2$
are constants, variances of $\textbf{x}$ and $\textbf{p}$ respectively,
constrained by the Heisenberg uncertainty relation $\Delta x^2 \Delta p^2 \geq
1/4$. This density matrix factorizes if and only if the terms proportional to
$p_1p_2$ in the exponent cancel, or equivalently when  the product $\Delta x^2
\Delta p^2$ has the smallest value allowed by the Heisenberg uncertainty
principle, i.e. for minimal wave packets \cite{PGG}. The factors are then
functions of $(p_s - p_i)^2 = 2m^2 + 2p_sp_i$, $i=1,2$ and the density matrix
can be written in the form (\ref{covcur})

Both approaches described here use simultaneously momenta and coordinates.
Their interpretations, however, are different. In the Wigner function, or
Shuryak, approach $\textbf{X}$ and $\textbf{K}$ are arithmetical averages of
the arguments of $\tilde{\rho}(\textbf{x},\textbf{x}')$ and
$\tilde{\rho}(\textbf{p},\textbf{p}')$ respectively. In the wave packets
approach they are labels of the sources, which may, but do not have to, be the
centres of the wave packets. Single particle energy is no problem since the
particles are on mass shell. Time in the wave packet formalism is just one more
label. In the Wigner function approach its meaning depends on the details of
the model.

\subsection{Difficulties with the final state interactions}

Final state interactions were further studied. It had been known
(cf. Section 4.5) that the two-body Coulomb interactions are
controlled by the dimensionless parameter $\xi_c = R/a_B$
\cite{PRA2}, where $a_B$ is the Bohr radius and $R$ is the radius
of the interaction region. For pion pairs $a_B \approx 386$ fm.
Consequently,  $\xi_c \approx 0.01$ for heavy ion collisions and
much less for the other cases. Therefore, it had been believed for
some time that Coulomb correction cannot be important. This
conclusion is not justified, because the factor $R$ in the small
parameter $\xi_c$ enters as an estimate of $1/|\textbf{q}|$. In
the very small $|\textbf{q}|$ region this is a gross
underestimate. Moreover, as seen from the Gamov factor, there are
large numerical coefficients. The leading correction in the small
$\xi_c$ limit is given by the Gamov factor \cite{GKW}. This
correction is significant and spoils the agreement with experiment
in the small $Q^2$ region (cf. e.g. \cite{GER}).

Pratt \cite{PRA2} has used the GGLP model with a Gaussian
distribution of sources and with the symmetrized product of plane
waves replaced by a symmetrized Coulomb wave functions. This
method of taking into account Coulomb interactions had been used
before \cite{KOO}, \cite{LEL}, but mainly in the study of the
Fermi-Dirac correlations among nucleons. Pratt found that
including the next term of the expansion in powers of $\eta$,
besides the Gamow factor, changes the estimated interaction radius
by about 20\% for $R$ as small as a few fermi.

In order to describe the two-body strong interaction among identical charged
pions it is enough to consider the $s$-wave, $I=2$ phase shift \cite{SUZ}. The
observed phase shift corresponds to repulsion, as one would expect for an
exotic two pion system where no resonances can be formed.  The key observation
\cite{BOW5} is that since the interactions are short range (about $0.2$ fm)
they are negligible for pions produced at a distance of one fermi, or so, from
each other. Consequently, this correction for strong interactions (not to be
confused with the corrections for resonance production) is negligible for the
study of Bose-Einstein correlation in heavy ion collisions. For $e^+e^-$
annihilations, on the other hand, this correlation can be important. According
to a rough estimate \cite{BOW5} it reduces the factor $\lambda$ by at least
30\%. Moreover, in agreement with experiment, it produces a minimum of the
correlation function at $Q \approx 0.6$ GeV, where the correlation function
drops below one. In spite of this success the quantitative applicability of
this model, where only an isolated pion pair is considered, is doubtful
\cite{BOW6}. For $\pi^+\pi^-$ pairs there is an attractive strong interaction
due to the $I=0$ $s$-wave phase shift. Therefore, three-body $\pi^+\pi^-\pi^+$
interactions give an effective attraction in the $\pi^+\pi^+$ system which,
according to rough estimates, could easily overcompensate the repulsion due to
the $I=2$ phase shift \cite{BOW6}. A more careful analysis including resonance
production \cite{BOW7} suggests that the strong interactions must somehow
cancel, because otherwise, the fully corrected $\lambda$ would come out
negative, which does not make sense. Another important remark \cite{BOW5} is
that experimentally, at low $Q^2$, the backgrounds formed from unlike sign
pions are not very different from the mixed backgrounds, where by construction
the strong interaction among members of pairs is absent. This supports the
point of view that final state interactions are less important than naively
expected. A study of two-body correlations for pairs of unlike sign pions could
bring here\footnote{And in fact latter brought.} interesting additional
information. Bowler's final conclusion for the strong final state interactions
is \cite{BOW8} "I think there is good reason to believe that in multiple
production the effects of final state interactions among all pairs largely
CANCEL and only the propagation and decay of resonances modifies the underlying
source structure". An exact analysis of such many body strong interaction is,
however, "impossibly complicated" \cite{BOW7}. Bowler concludes also
\cite{BOW6} that "the chaoticity and range parameters extracted from experiment
have no simple interpretation and should be regarded as purely descriptive".

\subsection{Early models with x - p correlations}

The improved formalism has been used to build models more
realistic than the early attempts of GGLP and KP. A common feature
of these models is that they include $x-\textbf{p}$ correlations.
Thus Pratt \cite{PRA1} considered

\begin{equation}\label{}
  S(x,\textbf{p}) = \delta(r - R)\delta(t)
  \exp\left[-\gamma\frac{E_p - v\hat{\textbf{r}}\cdot \textbf{p}}{T}\right],
\end{equation}
where the temperature $T$, velocity $v$ and Lorentz factor $\gamma
= 1/\sqrt{1 - v^2}$ are constant. The pions are ejected from a
sphere of radius $R$ and tend to go radially outwards. Thus,
contrary to the KP picture, there are $\textbf{x}- \textbf{p}$
correlations. Evaluating $R(\textbf{p}_1,\textbf{p}_2)$ for this
source function one finds \cite{PRA1} that the effective (i.e.
measured) radius of the interaction region is:

\begin{equation}\label{}
  R_{eff}(K) = R[(y\tanh y)^{-1} - (\sinh y)^2]^{1/2};\qquad y =
  \frac{|\textbf{K}|\gamma v}{T}.
\end{equation}
The expression in the square bracket equals $\frac{2}{3}$ for $|\textbf{K}|$ = 0 and
decreases towards zero with increasing $|\textbf{K}|$. However, in this model we know
that the true radius of the interaction region is R for all $\textbf{K}$. Let us
stress the implications of this important observation. One should study the
Bose-Einstein correlations at given $|\textbf{K}|$. A $|\textbf{K}|$ dependence of the
deduced radius of the interaction range implies the presence of
$\textbf{x}-\textbf{p}$ correlations. Wrong assumptions about the
$\textbf{x}-\textbf{p}$ correlations lead to wrong determinations of the radius of the
interaction region. For instance in Pratt's model averaging over $|\textbf{K}|$ and
interpreting the results as if there were no $\textbf{x}-\textbf{p}$ correlations one
obtains a radius below $\sqrt{2/3}$ of the correct (input) value.

A more realistic model with $x-\textbf{p}$  correlations, inspired by Bjorken's
inside-outside cascade, was given in ref. \cite{KOG} in the framework of the
covariant current approach. The authors chose

\begin{eqnarray}\label{kolgyu}
  D(x,p)&=& \frac{1}{\pi R_T^2}\delta(t - \tau_0 \cosh y)(\delta(z - \tau_0 \sinh y)
  e^{-\textbf{x}_T^2/R_T^2},\\
  j(\frac{p_s p}{m}) &=& \sqrt{a}\exp\left[-\frac{p_sp}{2mT}\right],
\end{eqnarray}
where $a, T, R_T$ and  $\tau_0$ are constants. This corresponds to
a distribution of sources which is uniform in rapidity and
concentrated at the longitudinal proper time $\tau_0 \equiv
\sqrt{t^2 - z^2}$. Each source in its rest frame produces pions
according to a "pseudothermal model" which is a modification of
the Boltzmann distribution chosen so as to make the subsequent
integrations possible to perform analytically. The corresponding
single particle pseudo density matrix is \cite{KOG}, \cite{PGG}

\begin{eqnarray}\label{}
\rho(\textbf{p}_1,\textbf{p}_2)& = &a K_0(\sqrt{u})e^{-q_t^2R^2_T};\\
 u& = &\left[\frac{m_{1T}+m_{2T}}{2T} - i\tau_0(m_{1T} - m_{2T})\right]^2 +\nonumber
 \\& &
 2\left(\frac{1}{4T^2} + \tau_0^2\right)m_{1T}m_{2T}\left[\cosh(y_1 - y_2) - 1\right],
\end{eqnarray}
where $m_{iT}$ are the transverse masses, $y_i$ the rapidities of the two
particles and $K_0$ is the modified Bessel function. It is interesting that
this model, extended by the inclusion of resonances which requires the use of a
Monte Carlo program, explains the data of the NA35 experiment \cite{BAM}, which
had been previously explained \cite{BGT} by assuming the presence of a
quark-gluon plasma. Resonance production is the most important correction,
because other corrections like a spread in the freeze-out proper time $\tau_0$
or a nonuniform distribution in rapidity tend to cancel \cite{PAG}. This
analysis can be generalized and made more rigorous \cite{PGG}.  Let us also
note attempts to use Monte Carlo programs to get more realistic distributions
of the sources \cite{HUM1}, \cite{CSO1}.

A rather different approach to Bose-Einstein correlations in hadron and heavy
ion collisions was developed by Makhlin, Sinyukov and collaborators (cf.
\cite{MAS1}, \cite{AMS}, \cite{MAS2}, \cite{SIN1} and references quoted there).
Their starting point was Landau's hydrodynamic model. According to this model
the production of hadrons in high energy hadronic interactions proceeds in
three stages. First the content of the interaction region thermalizes. This
happens very fast. Then the fluid expands and cools according to the laws of
hydrodynamics. Whenever the temperature of an element of the fluid drops to a
critical temperature $T = T_{cr}$, its content gets converted into hadrons
which have the thermal equilibrium distribution corresponding to temperature
$T_{cr}$. This hadronization of a fluid element is also assumed to be a very
rapid process. Since at every space point the temperature is a function of
time, the space-time points where $T(x,t) = T_{cr}$ form a three-dimensional
hypersurface $\Sigma_{cr}$ in space-time and all the hadrons are produced
there. In order to get predictions one should in principle assume some initial
conditions i.e. the distribution at the beginning of the hydrodynamic stage,
solve the hydrodynamical equations with these initial conditions and find the
hypersurface $\Sigma_{cr}$ as well as the velocity distribution of the fluid on
it. This would be a very difficult task, but there are simple estimates. E.g.
one can define $\Sigma_{cr}$ by the equation $t^2 - z^2 = \tau^2$, where $\tau$
is a constant and assume that the fluid velocity is parallel to the $z$ axis
and at space-time point $\textbf{x},t$ equal $z/t$. The authors suggest that by
selecting events with a given heights of the central plateau in rapidity one
can approximately fix $\tau$. There is also a more complicated estimate due to
Landau. When transverse expansion is included, parts of the hypersurface
$\Sigma_{cr}$ become time like which complicates the formalism, but can be
dealt with \cite{SIN1}. The hydrodynamic approach got very popular later,
though usually it just means introducing collective velocities of the sources.

\subsection{General proposals}

Some results more general than specific models have also been given. For heavy ion
collisions the plausible assumption that the transverse profile of the interaction
region is given by the overlap of the two nuclear distributions explains easily two
experimental facts. For a given pair of ions the average transverse radius of the
interaction region $R_T$ increases with decreasing impact parameter \cite{BARS} and
for central collisions $R_T$ is given by the radius of the smaller of the two
colliding ions \cite{CAK}.

In ref. \cite{KML} the authors, generalizing earlier results from \cite{POC}, proved
that when $\rho(\textbf{p}_1,\textbf{p}_2)$ integrated over
$\textbf{p}_1+\textbf{p}_2$ can be written in the form $\rho(a^2\textbf{q}_T^2 +
q_z^2)$, where $a$ is a constant, the distribution of the angle $\theta$ between
vector $\textbf{q}$ and the axis of the event $z$ is

\begin{equation}\label{podcze}
\phi(\cos\theta) = \frac{a^2}{2[a^2 + (1-a^2)\cos^2\theta]^{3/2}}.
\end{equation}
This formula, which follows by simple integration, can be useful to distinguish, cigar
shaped ($a^2 < 1$) from pancake shaped ($a^2 > 1$) sources. Pratt \cite{PRA3} and
Bertsch \cite{BER} reintroduced for the difference of momenta $\textbf{q}$ the
coordinate system which had been used some ten years before by Grassberger
\cite{GRAS}. The notation and terminology is from Bertsch \cite{BER}: $q_L$ for the
longitudinal component along the event axis, $q_o$ for the outward component parallel
to $\textbf{K}_T$ and $q_s$ for the sideward component perpendicular to both the event
axis and to $\textbf{K}$. His formula\footnote{Actually in the paper the coefficient
of $q_L^2$ reads $\frac{1}{2}R_L^{-2}$, but this seems to be a misprint.}

\begin{equation}\label{}
  R(p_1,p_2) = e^{-(R_s^2q_s^2 + R_0^2q_o^2 + R_L^2q_L^2)}
\end{equation}
soon became very popular. The main point of both authors was that, if the time
interval when the pions are produced is long - this could be due to a slowly
hadronizing quark-gluon plasma but other possibilities were also considered - then one
expects $R_o \gg R_s$, because the pions produced late behave like pions produced far
along the $out$ direction. Let us finally note a diagrammatic classification of the
various terms occurring in the theory of multiparticle Bose-Einstein correlation with
coherence included \cite{BIY2}. For an attempt to estimate the effect of multibody
Bose-Einstein correlations on the two particle correlation function by direct
calculation of the permanents see \cite{ZAJ1}.

Podgoretsky \cite{POD1} \cite{POD2} studied the dependence of the parameters
$R$ and $\tau$ on the boosts along the event axis. Let us denote the event axis
by $z$ and the axes in the $out$ and $side$ directions by $x$ and $y$
respectively. Let us further assume that in some reference frame the
distribution of sources is given by a formula similar to the Yano-Koonin
formula (\ref{cmsyak}):

\begin{equation}\label{}
\rho(x) = \frac{1}{4\pi^2R^2\tilde{R}T}\exp\left[-\frac{x^2 + y^2}{2R^2} -
\frac{z^2}{2\tilde{R}^2} - \frac{\tau^2}{2T^2}\right].
\end{equation}
Then, in the reference frame moving with respect to this frame with velocity
$\beta$ (Lorentz factor $\gamma$) parallel to the $z$ axis, the distribution of
$\textbf{q}$ is \cite{POD1}

\begin{equation}\label{podpar}
  W \sim 1 + \exp\left[-q_x^2R^2 - q_y^2R^2 - \gamma^2\{q_z(1-\beta u_z) - q_x\beta u_x\}^2\tilde{R}^2 -
  \gamma^2\{q_z(\beta - u_z) - q_xu_x\}^2T^2\right].
\end{equation}
Here $\textbf{u}$ is the velocity of the pion pair $\textbf{K}/K_0$ and the
identity

\begin{equation}\label{}
  q_0 = \textbf{u}\cdot \textbf{K}
\end{equation}
has been used. Yano and Koonin \cite{YAK} got a similar result, but instead of
using explicitly the Lorentz transformation they expressed the exponent in
terms of invariants, see formula (\ref{yankoo}). Podgoretsky noticed that
whenever there is a frame where all the pion sources are at rest and where
their distribution is invariant under the change of orientation of the event
axis, the parameters $R_L$ and $\tau$ must be extremal in this frame. Thus, it
could be possible to identify the frame, where all the sources are at rest.
Early estimates \cite{AAG1}, \cite{AAG2} favored the quark frame, but later the
much more detailed analysis of \cite{AGB0} demonstrated, that there is no clear
minimum of $R_L$ as a function of $p_p/p_{\pi^+}$ and consequently no
privileged frame, where all the sources are at rest. On the other hand the
parameterization (\ref{podpar}) turned out to be very useful.

\section{Conclusions}

In his review talk entitled \textit{The GGLP effect alias Bose-Einstein effect
alias HBT effect} at a conference in Marburg in 1990 \cite{GOL2} G. Goldhaber
said: \textit{What is clear is that we have been working on this effect for
thirty years. What is not as clear is that we have come much closer to a
precise understanding of the effect}. Let us discuss this statement from the
present point of view.

There is no doubt that GGLP was a great paper. It correctly identified the
Bose-Einstein correlations as the reason for the small $Q^2$ enhancement for
the like sign pion pairs in $\overline{p}p$ interactions\footnote{Amusingly, in
the special case of $\overline{p}N$ interactions at low energy there is now
evidence \cite{GAS}, \cite{AAN}, \cite{APO} that the GGLP effect is due to
resonances more than to BEC between directly produced pions}. Moreover, the
authors proposed the invariant Gaussian parameterization, noticed the
possibility of getting from BEC information about the interaction region,
pointed out that a joint analysis of many experimental distributions should be
made and mentioned the possible importance of resonances.

The most important message: the possibility of measuring the interaction
region, however, passed unnoticed. Only after the work of  Kopylov,
Podgoretsky, and Cocconi in the early seventies it became gradually accepted
that BEC are more interesting as a tool to study the interaction region than as
a peculiarity of the distribution of the relative momentum of identical pions.
This was certainly a breakthrough. Another breakthrough happened on the
technical side. The GGLP method of modifying the integrand of the integral over
phase space was applicable only to low multiplicity exclusive channels, while
the method, pioneered by Kopylov and Podgoretsky, of modifying the correlation
function could be applied at any (sufficiently high) multiplicity and also to
inclusive distributions (at sufficiently high energy).

The relation between the properties of the interaction region and the data on
interparticle correlations turned out to be much more complicated than
anticipated. Final state interactions, resonances, experimental resolution and
particle identification are all important factors. Even, however, when all
these technical problems are solved, important sources of systematic error
remain. Once it is recognized that the hadrons are not all produced instantly
and simultaneously, the problem arises: how to reconstruct a four-dimensional,
space-time distribution of sources from the measured three-dimensional
distribution of momenta. Moreover, BEC can be only observed among particles
with similar momenta. Particles with similar momenta, according to many models,
are produced in a homogeneity region, which is a fraction of the total
interaction region. Thus, what one measures is the size of the homogeneity
region and not that of the total hadronization region. In other words, it is
not possible to draw conclusions from BEC without previous knowledge about the
position- (in space-time) momentum correlations of the produced hadrons.

The conclusion is that attempts to get information about the interaction region
from a model independent analysis of the data seem hopeless. It is necessary to
start with a model and then, within the model, it may be possible to use the
data on BEC to get information about the interaction region. This information
may be crucial, like in the case of hydrodynamic models, or of marginal
interest like in string models.

To summarize: the idea to study BEC as a way of learning about the interaction
region has certainly made their study much more interesting. Many problems not
anticipated by GGLP have been identified and some of them have been solved. A
reliable and non controversial method of getting quantitative information about
the interaction region from BEC is still, however, not available, as was
strikingly illustrated by the failure of the predictions made for RHIC. If this
is meant by \textit{the precise understanding of the effect} Goldhaber's
position is still defendable.

\end{document}